\documentclass[pdflatex,sn-mathphys-num,iicol]{sn-jnl}

\usepackage{amsmath}
\usepackage{amsfonts}
\usepackage{graphicx}
\usepackage{subfigure}
\usepackage{multirow}
\usepackage{booktabs}
\usepackage{adjustbox}
\usepackage{xcolor}
\usepackage{microtype}

\hypersetup{colorlinks=true,citecolor=blue,urlcolor=blue,linkcolor=blue}

\begin{document}

\title[aQGC limits via $ZZ\gamma$ production at FCC-hh]{Probing the limits on anomalous quartic gauge couplings via $ZZ\gamma$ production in the $\ell \ell \nu \nu \gamma$ channel at FCC-hh}

\author*[1]{\fnm{A.} \sur{Yilmaz}}\email{ali.yilmaz@ibu.edu.tr}

\author[2]{\fnm{A.} \sur{Senol}}\email{senol\_a@ibu.edu.tr}

\author[2]{\fnm{H.} \sur{Denizli}}\email{denizli\_h@ibu.edu.tr}

\author[3]{\fnm{I.} \sur{Turk Cakir}}\email{iturk@ankara.edu.tr}

\author[4]{\fnm{O.} \sur{Cakir}}\email{ocakir@science.ankara.edu.tr}

\affil*[1]{\orgdiv{Department of Computer Engineering}, \orgname{Bolu Abant Izzet Baysal University}, \orgaddress{\postcode{14030}, \city{Bolu}, \country{Turkey}}}

\affil[2]{\orgdiv{Department of Physics}, \orgname{Bolu Abant Izzet Baysal University}, \orgaddress{\postcode{14030}, \city{Bolu}, \country{Turkey}}}

\affil[3]{\orgdiv{Institute of Accelerator Technologies}, \orgname{Ankara University}, \orgaddress{\postcode{06100}, \city{Ankara}, \country{Turkey}}}

\affil[4]{\orgdiv{Department of Physics}, \orgname{Ankara University}, \orgaddress{\postcode{068300}, \city{Ankara}, \country{Turkey}}}

\abstract{
In this study, the sensitivity to anomalous quartic gauge couplings (aQGCs) is projected via $pp \rightarrow ZZ\gamma$ production in the 100~TeV proton-proton Future Circular Collider – hadron-hadron (FCC-hh) for an integrated luminosity of 30~ab$^{-1}$. The $\ell\ell\nu\nu\gamma$ final state under consideration consists of a same-flavor, opposite-sign lepton pair (electrons or muons) from one $Z$ boson, the invisible decay of the other $Z$ boson into neutrinos, and an accompanying photon. The FCC-hh detector response and its effects on the reconstructed objects are included through a realistic detector simulation. Three multivariate techniques are employed to separate the signal from the relevant SM backgrounds. Unitarity is preserved by a strict, operator-dependent bound on the total transverse mass ($M_T^{tot}$) of the system. The median expected significances are calculated within the Asimov approximation for one anomalous coupling varied at a time and for background systematic uncertainties between 0\% and 10\%. The highest separation power is obtained with the deep neural network method. The resulting 95\% confidence level limits on $|f_{T0}/\Lambda^{4}|$, $|f_{T8}/\Lambda^{4}|$, $|f_{T9}/\Lambda^{4}|$ and $|f_{M2}/\Lambda^{4}|$ in the combined $e+\mu$ channel without systematic uncertainties are $2.83\times 10^{-3}$, $1.65\times 10^{-3}$, $3.81\times 10^{-3}$ and $8.97\times 10^{-3}$~TeV$^{-4}$, respectively. 
We have an order of magnitude improvement when compared to current LHC limits with the assumption of 5\% systematic uncertainty.}


\maketitle


\section{Introduction}
\label{sec:intro}

The Standard Model (SM) describes particle interactions accurately up to the electroweak scale. The Higgs boson discovery in 2012 at the Large Hadron Collider (LHC)~\cite{ATLAS:2012yve,CMS:2012qbp} completed the SM particle content and confirmed electroweak symmetry breaking. Despite its success, the SM is essentially an effective theory. Open questions such as dark matter, neutrino masses, and the baryon asymmetry strongly suggest new physics (NP) awaits at higher energies~\cite{Martin:1997ns,Bertone:2004pz}.
Measuring gauge boson self-interactions provides a direct test of the electroweak sector, offering an indirect probe for NP. The SM gauge symmetry tightly constrains these couplings. While charged triple gauge couplings (TGCs) such as $W^+W^-\gamma$ and $W^+W^-Z$ are well-measured, neutral TGCs ($ZZ\gamma$, $Z\gamma\gamma$) vanish at tree level in the SM~\cite{Hagiwara:1986vm,Gounaris:1999kf}. Quartic gauge couplings (QGCs) are similarly constrained. Any observed deviation here would be a clear signal of NP.\\
Multi-boson production at colliders is the standard way of probing these interactions, and triboson final states provide direct access to QGCs. The $pp \rightarrow ZZ\gamma$ process involves $ZZ\gamma\gamma$ and $ZZZ\gamma$ vertices, making it particularly sensitive to anomalous couplings~\cite{Eboli:2006wa,Eboli:2016kko}. While LHC studies of this process have limited sensitivity due to small cross-sections, CMS recently found evidence for $ZZ\gamma$ (in the $4\ell\gamma$ final state)~\cite{CMS:2026vhr} and $WZ\gamma$ production~\cite{CMS:2025oey} at 13~TeV. These results show that triboson channels involving a photon are viable for anomalous QGC (aQGC) searches.
The increase to a 100~TeV collision energy at the proposed Future Circular Collider – hadron-hadron (FCC-hh) will provide the large event rates needed to extend the reach of rare SM processes and NP searches~\cite{FCC:2018byv,FCC:2018vvp}. The $\ell\ell\nu\nu\gamma$ final state of $ZZ\gamma$ ($Z \rightarrow \ell^+\ell^-$ and $Z \rightarrow \nu\bar{\nu}$) is chosen because it provides an optimal trade-off between signal yield and background suppression. It avoids the large QCD background of hadronic decays while retaining a higher branching ratio than the fully leptonic state.\\
Effective Field Theory (EFT) parametrizes deviations in self-interactions by adding higher-dimensional operators to the SM Lagrangian~\cite{Degrande:2012wf,Brivio:2017vri}. For neutral gauge self-interactions, the leading NP effects appear at dimension-8 (dim-8). In this study, the $T$ class (built purely from field strength tensors) and the $M$ class (which include the Higgs covariant derivative) operators are considered~\cite{Eboli:2006wa,Eboli:2016kko}. Specifically, the coefficients $f_{T0}/\Lambda^4$, $f_{T8}/\Lambda^4$, $f_{T9}/\Lambda^4$, and $f_{M2}/\Lambda^4$ are studied, with one coupling varied at a time.\\
Current aQGC constraints come from LHC multi-boson and vector boson scattering (VBS) measurements. The most stringent limits on $f_{T8}/\Lambda^4$ and $f_{T9}/\Lambda^4$ are obtained from the ATLAS measurement of the electroweak $Z(\nu\bar{\nu})\gamma jj$ production at $\sqrt{s}=13$~TeV~\cite{ATLAS:2022nru}, while the tightest bounds on $f_{T0}/\Lambda^4$ and $f_{M2}/\Lambda^4$ come from the CMS analysis of VBS in the semileptonic final states~\cite{CMS:2025dbm}. The electroweak diboson production in association with a high-mass dijet system has also been measured by ATLAS in the semileptonic channels~\cite{ATLAS:2025omi}. Limits on aQGCs (dim-8) have also been set through electroweak $W\gamma jj$~\cite{CMS:2016gct}, $Z\gamma$ and $Z\gamma\gamma$~\cite{ATLAS:2016qjc,CMS:2021jji,ATLAS:2022wmu}, $W^{+}W^{-}\gamma$~\cite{ATLAS:2025yxf}, and same-sign $WW$ scattering~\cite{ATLAS:2016snd,CMS:2017fhs,ATLAS:2019cbr} measurements at the LHC. Recent phenomenological studies have also explored the potential of future colliders to improve the constraints on the neutral gauge boson couplings, including anomalous neutral triple gauge boson interactions via $ZZ$ and $ZZ\gamma$/$Z\gamma\gamma$ production at the FCC-hh and the HL-/HE-LHC~\cite{Senol:2018cks,Yilmaz:2019cue,Senol:2019qyl,Yilmaz:2021ule,Senol:2019swu} and neutral quartic gauge couplings at a muon collider~\cite{Gutierrez-Rodriguez:2025wcy}.

In this paper, the FCC-hh sensitivity to dim-8 operators in $ZZ\gamma$ production is projected via the $\ell\ell\nu\nu\gamma$ channel. The constraints on $f_{T0}/\Lambda^4$, $f_{T8}/\Lambda^4$, $f_{T9}/\Lambda^4$, and $f_{M2}/\Lambda^4$ are calculated using three different multivariate techniques, namely Boosted Decision Trees (BDT),  Boosted Decision Trees Decorrelated (BDTD), and Deep Neural Networks (DNN), across the electron, muon, and combined $e+\mu$ channels. The paper is organized as follows: Section~\ref{sec:theory} describes the theoretical framework, Section~\ref{sec:generation} details the event generation, and Section~\ref{sec:selection} outlines the event selection and multivariate analysis. Results are discussed in Section~\ref{sec:results}, followed by conclusions in Section~\ref{sec:conclusion}.

\section{Theoretical framework}
\label{sec:theory}

The SM electroweak gauge boson self-interactions are fixed by the non-Abelian $SU(2)_L \times U(1)_Y$ symmetry. The kinetic terms and self-interactions are:

\begin{equation}
\mathcal{L}_{\text{gauge}} = -\frac{1}{4}W^a_{\mu\nu}W^{a\mu\nu} - \frac{1}{4}B_{\mu\nu}B^{\mu\nu}
\end{equation}
where $W^a_{\mu\nu}$ and $B_{\mu\nu}$ are the field strength tensors associated with the $SU(2)_L$ and $U(1)_Y$ gauge fields, respectively:

\begin{equation}
W^a_{\mu\nu} = \partial_\mu W^a_\nu - \partial_\nu W^a_\mu + g\epsilon^{abc}W^b_\mu W^c_\nu
\end{equation}
\begin{equation}
B_{\mu\nu} = \partial_\mu B_\nu - \partial_\nu B_\mu
\end{equation}
here, $g$ is the $SU(2)_L$ coupling constant, and $\epsilon^{abc}$ are the structure constants of the $SU(2)$ group. After electroweak symmetry breaking, the physical gauge bosons $W^\pm$, $Z$, and $\gamma$ emerge as linear combinations of the gauge eigenstates, and their self-interactions are determined by the above terms.\\
The quartic gauge couplings in the SM involve vertices with four gauge bosons, such as $W^+W^-W^+W^-$, $W^+W^-ZZ$, $W^+W^-Z\gamma$, and $W^+W^-\gamma\gamma$. However, neutral QGCs involving only neutral gauge bosons, such as $ZZZ\gamma$ or $ZZ\gamma\gamma$, are not present at tree level in the SM~\cite{Hagiwara:1986vm,Gounaris:1999kf}. These neutral QGCs only appear at loop level and are strongly suppressed. This makes them clean targets for new physics searches.\\
EFT parametrizes deviations from the SM by adding higher-dimensional operators to the Lagrangian:

\begin{equation}
\mathcal{L}_{\text{eff}} = \mathcal{L}_{\text{SM}} + \sum_{d>4} \sum_i \frac{f_i^{(d)}}{\Lambda^{d-4}} \mathcal{O}_i^{(d)}
\end{equation}
where $\mathcal{O}_i^{(d)}$ are operators of dimension $d$, and $f_i^{(d)}$ are dimensionless coefficients that encode the strength of NP contributions.\\
For neutral gauge self-interactions, the leading NP effects only begin at dim-8~\cite{Degrande:2012wf,Eboli:2016kko}. Operators from the $T$ class and the $M$ class, as classified in Ref.~\cite{Eboli:2006wa}, are considered:

\begin{equation}
\mathcal{O}_{T,0} = \text{Tr}[W_{\mu\nu}W^{\mu\nu}] \times \text{Tr}[W_{\alpha\beta}W^{\alpha\beta}]
\end{equation}
\begin{equation}
\mathcal{O}_{T,8} = B_{\mu\nu}B^{\mu\nu} \times B_{\alpha\beta}B^{\alpha\beta}
\end{equation}
\begin{equation}
\mathcal{O}_{T,9} = B_{\alpha\mu}B^{\mu\beta} \times B_{\beta\nu}B^{\nu\alpha}
\end{equation}
\begin{equation}
\mathcal{O}_{M,2} = \left[B_{\mu\nu}B^{\mu\nu}\right] \times \left[(D_{\beta}\Phi)^{\dagger} D^{\beta}\Phi\right]
\end{equation}
{\sloppy
Operators $\mathcal{O}_{T,8}$ and $\mathcal{O}_{T,9}$ are particularly relevant for $ZZ\gamma$ production because they contain the $B$ field strength tensor, which projects onto both the $Z$ and $\gamma$ fields. These operators only induce neutral quartic vertices~\cite{Eboli:2016kko,Gounaris:1999kf}. The relevant effective Lagrangian terms are:
\par}

\begin{equation}
\mathcal{L}_{\text{eff}} \supset \frac{f_{T0}}{\Lambda^4}\mathcal{O}_{T,0} + \frac{f_{T8}}{\Lambda^4}\mathcal{O}_{T,8} + \frac{f_{T9}}{\Lambda^4}\mathcal{O}_{T,9} + \frac{f_{M2}}{\Lambda^4}\mathcal{O}_{M,2}
\end{equation}
where the goal of this analysis is to constrain the dimensionless coefficients $f_{T0}/\Lambda^4$, $f_{T8}/\Lambda^4$, $f_{T9}/\Lambda^4$, and $f_{M2}/\Lambda^4$.
Table~\ref{tab:operator_vertex_map} shows the vertices modified by each operator. All four affect $ZZZ\gamma$ and $ZZ\gamma\gamma$. However, $\mathcal{O}_{T,0}$ broadly modifies all quartic vertices, whereas $\mathcal{O}_{M,2}$ leaves $Z\gamma\gamma\gamma$ and $\gamma\gamma\gamma\gamma$ untouched.

\begin{table*}[htbp]
\centering
\caption{Quartic gauge-boson vertices modified by the dim-8 operators following Ref.~\cite{Eboli:2006wa}. A quartic vertex
modified by a given operator is marked with X, and with O otherwise. The vertices
$ZZZ\gamma$ and $ZZ\gamma\gamma$ (boldface), which drive the $pp\to ZZ\gamma$
process, receive contributions from all four operators.}
\label{tab:operator_vertex_map}
\begin{tabular}{l c c c c c c c c c}
\toprule
 & $WWWW$ & $WWZZ$ & $ZZZZ$ & $WW\gamma Z$ & $WW\gamma\gamma$ &
 $\boldsymbol{ZZZ\gamma}$ & $\boldsymbol{ZZ\gamma\gamma}$ &
 $Z\gamma\gamma\gamma$ & $\gamma\gamma\gamma\gamma$ \\
\midrule
$\mathcal{O}_{T,0}$ & X & X & X & X & X & X & X & X & X \\
$\mathcal{O}_{T,8}$ & O & O & X & O & O & X & X & X & X \\
$\mathcal{O}_{T,9}$ & O & O & X & O & O & X & X & X & X \\
$\mathcal{O}_{M,2}$ & O & X & X & X & X & X & X & O & O \\
\bottomrule
\end{tabular}
\end{table*}

\subsection{The $pp \rightarrow ZZ\gamma$ process}

\begin{figure}[htbp]
\centering
\adjustbox{valign=c}{\includegraphics[scale=0.85]{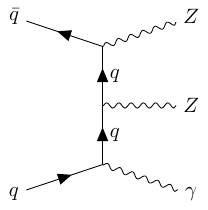}} 
\adjustbox{valign=c}{\includegraphics[scale=0.85]{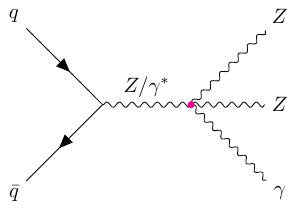}}
\caption{Representative Feynman diagrams for the $pp \rightarrow ZZ\gamma$ process. The SM production is shown on the left and the production including the quartic gauge boson vertex sensitive to the anomalous contributions on the right.}
\label{fig:feynman}
\end{figure}

In the SM, leading order (LO) $ZZ\gamma$ production proceeds through $q\bar{q} \rightarrow ZZ\gamma$~\cite{Bozzi:2009ig}. A representative diagram is shown in the left panel of Figure~\ref{fig:feynman}, where the two $Z$ bosons and the photon are radiated from the initial-state quark line through $t$- and $u$-channel quark exchange. The right panel shows the anomalous production, in which an $s$-channel $Z/\gamma^{*}$ couples to the $ZZ\gamma$ final state through a quartic gauge boson vertex.

Introducing anomalous QGCs allows for diagrams containing $ZZ\gamma\gamma$ and $ZZZ\gamma$ vertices. Because the amplitudes for these interactions grow with energy, they eventually violate unitarity. Rather than being damped with an arbitrary dipole form factor, they are restricted by a strict, operator-dependent bound derived from partial-wave unitarity~\cite{Rauch:2016pai,Covarelli:2021gyz}. This bound is imposed at the analysis level, as described in Sect.~\ref{sec:selection}.

With one coupling varied at a time, the cross section is parameterized as:

\begin{equation}
\sigma\!\left(f/\Lambda^4\right) = \sigma_{SM} + \sigma_{int}\,\frac{f}{\Lambda^4} + \sigma_{quad}\left(\frac{f}{\Lambda^4}\right)^2
\label{eq:xsec_param}
\end{equation}
where $\sigma_{SM}$ is the SM cross-section, $\sigma_{int}$ represents the interference between the SM and the anomalous contribution, and $\sigma_{quad}$ corresponds to the pure anomalous term.
The simulations include up to one anomalous vertex insertion per event, so the generated samples contain the SM--anomalous interference along with the pure anomalous contribution. The signal is defined as the excess over the SM yield, $S = N(f) - N_{SM}$. The constant SM term of Eq.~(\ref{eq:xsec_param}) cancels in this difference, so that the excess is governed by the interference and the pure anomalous terms alone,

\begin{equation}
S \propto \sigma_{int}\,\frac{f}{\Lambda^{4}} + \sigma_{quad}\left(\frac{f}{\Lambda^{4}}\right)^{2}.
\label{eq:excess}
\end{equation}
At the couplings scanned here the quadratic term dominates over the interference. 
This scaling is verified on the simulated samples in Sect.~\ref{sec:results}.
A 100~TeV collision energy at the FCC-hh enhances $ZZ\gamma$ cross-sections. Because anomalous amplitudes grow with energy, this provides a kinematic advantage to set constraints beyond the reach of the LHC.
The primary backgrounds for the $\ell\ell\nu\nu\gamma$ final state are $WW\gamma$, $WZ\gamma$, $t\bar{t}\gamma$, $Z\gamma\gamma$ and $ZZZ$.

\section{Generation of signal and background events}
\label{sec:generation}

{\sloppy
The signal and background event samples are generated at leading order with \texttt{MadGraph5\_aMC@NLO} \texttt{v3.5.7}~\cite{Alwall:2014hca}, where the dim-8 operators are implemented through the FeynRules~\cite{Alloul:2013bka} UFO~\cite{Degrande:2011ua} models \texttt{SM\_LT012\_UFO} for $\mathcal{O}_{T,0}$, \texttt{SM\_LT8\_LT9\_UFO} for $\mathcal{O}_{T,8}$ and $\mathcal{O}_{T,9}$, and \texttt{SM\_LM0123\_UFO} for $\mathcal{O}_{M,2}$, together with the default \texttt{NNPDF2.3LO} parton distribution functions~\cite{Ball:2012cx}. The signal process is generated as $pp \rightarrow ZZ\gamma$ with up to one anomalous vertex insertion, and the SM sample is generated with the anomalous couplings set to zero. The   $WW\gamma$, $WZ\gamma$, $t\bar{t}\gamma$, $Z\gamma\gamma$, and $ZZZ$ background samples are produced with the same generator setup. Parton showering and hadronization are performed with \texttt{PYTHIA~v8.316}~\cite{Sjostrand:2014zea}. The detector response is simulated with \texttt{Delphes~v3.5.1}~\cite{deFavereau:2013fsa} using the official FCC-hh baseline detector card \texttt{FCChh.tcl} distributed with \texttt{Delphes}, without any modification of its parametrization.
\par}

\begin{figure}[htbp]
\centering
\includegraphics[width=0.98\columnwidth]{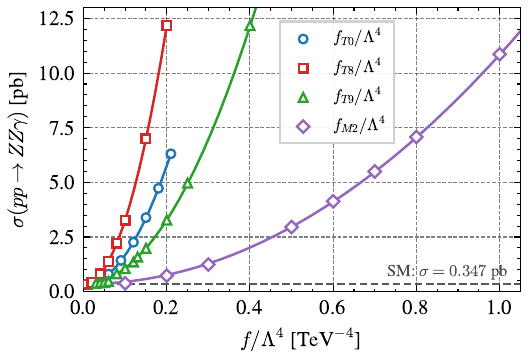}
\caption{Generator-level cross section of the $pp\to ZZ\gamma$ process as a function of the anomalous coupling $f/\Lambda^{4}$ (one operator varied at a time, the others set to zero), in units of TeV$^{-4}$, for the $f_{T0}/\Lambda^{4}$, $f_{T8}/\Lambda^{4}$, $f_{T9}/\Lambda^{4}$ and $f_{M2}/\Lambda^{4}$ operators. Markers are the simulated points and the solid curves the quadratic fits of Eq.~(\ref{eq:xsec_param}). The dashed line indicates the SM cross section $\sigma_{SM}=0.347$~pb.}
\label{fig:scaling}
\end{figure}

We generated signal samples by varying one coupling at a time. The scanned ranges are $0.03\mbox{--}0.21$ for $f_{T0}/\Lambda^{4}$, $0.02\mbox{--}0.20$ for $f_{T8}/\Lambda^{4}$, $0.03\mbox{--}1.00$ for $f_{T9}/\Lambda^{4}$ and $0.1\mbox{--}3.0$ for $f_{M2}/\Lambda^{4}$, in units of TeV$^{-4}$. For each signal coupling point, the SM sample, and each background process, $3\times10^{6}$ events are generated per channel.
Figure~\ref{fig:scaling} shows the generator-level cross section versus coupling strength. It matches the SM value (0.347 pb) at zero coupling and grows quadratically. The $f_{T8}/\Lambda^{4}$ operator yields the largest cross section increase, while $f_{M2}/\Lambda^{4}$ yields the smallest.

The final state is reconstructed from a same-flavor opposite-sign lepton pair, a photon, and missing transverse energy, as detailed in Sect.~\ref{sec:selection}.

\section{Event selection and multivariate analysis}
\label{sec:selection}

The preselection criteria are written in the Analysis Description Language and applied to the simulated signal and background samples with \texttt{CutLang~v2.14.1}~\cite{Unel:2021edl}. The invisible $Z$ decay is reconstructed from the missing transverse energy. The preselection requires at least one photon, at least one same-flavor opposite-sign lepton pair, and the leading lepton pair's invariant mass must fall within 5~GeV of the nominal $Z$ mass. The resulting event selection flow is summarized in Table~\ref{tab:cutflow_preML}.

\begin{table*}[htbp]
\centering
\caption{Event selection cut flow applied by CutLang~\cite{Unel:2021edl} to the aQGC signal ($ZZ\gamma$ with $f_{T8}/\Lambda^4=0.02$~TeV$^{-4}$) and to the SM background samples. Event counts are given for the muon channel, with the electron channel in parentheses. $N_{\ell}$ and $N_{\gamma}$ denote the numbers of selected leptons and photons, and $m_{\ell\ell}$ the invariant mass of the leading lepton pair. Each sample is generated with $3\times10^{6}$ events. The counts are raw simulated event numbers, not normalized to the integrated luminosity, and therefore quantify the selection efficiency within each sample rather than the expected event rates. The luminosity-normalized yields after the multivariate selection and the unitarity bound are given in Table~\ref{tab:yields_wp}. The same selection is applied to all signal and background samples; the surviving events are passed to the multivariate analysis.}
\label{tab:cutflow_preML}
\begin{adjustbox}{max width=\textwidth}
\begin{tabular}{l ccccccc}
\toprule
Cut & $ZZ\gamma$ (aQGC) & $ZZ\gamma$ & $WW\gamma$ & $WZ\gamma$ & $t\bar{t}\gamma$ & $Z\gamma\gamma$ & $ZZZ$ \\
\midrule
Total events & \multicolumn{1}{c}{3000000} & \multicolumn{1}{c}{3000000} & \multicolumn{1}{c}{3000000} & \multicolumn{1}{c}{3000000} & \multicolumn{1}{c}{3000000} & \multicolumn{1}{c}{3000000} & \multicolumn{1}{c}{3000000} \\
$N_{\ell}\ge 2$, $N_{\gamma}\ge 1$, $E_T^{miss}>0$ & 79124 (60193) & 65555 (46746) & 14687 (10238) & 35594 (25931) & 15592 (10058) & 55339 (43442) & 4542 (2930) \\
Opposite-sign same-flavour pair & 78300 (59468) & 64634 (45952) & 14633 (10173) & 33584 (24078) & 15384 (9916) & 55031 (43152) & 4385 (2546) \\
$|m_{\ell\ell}-m_Z|<5$ GeV & 66895 (53047) & 59529 (43426) & 763 (348) & 30055 (21712) & 572 (421) & 48809 (34238) & 2683 (1320) \\
\bottomrule
\end{tabular}
\end{adjustbox}
\end{table*}

The total transverse mass of the $Z\gamma + E_T^{miss}$ system is defined as
\begin{equation}
\begin{split}
M_T^{tot} = \big[\, & 2\,p_T^{\ell\ell}\,E_T^{miss}\,(1-\cos\Delta\phi_{\ell\ell,\nu}) \\
& + 2\,p_T^{\ell\ell}\,p_T^{\gamma}\,(1-\cos\Delta\phi_{\ell\ell,\gamma}) \\
& + 2\,p_T^{\gamma}\,E_T^{miss}\,(1-\cos\Delta\phi_{\gamma,\nu}) \,\big]^{1/2},
\end{split}
\label{eq:mttot}
\end{equation}
where $p_T^{\ell\ell}$ is the transverse momentum of the reconstructed leptonic $Z$ boson, $p_T^{\gamma}$ that of the photon, $E_T^{miss}$ the missing transverse energy carried by the invisible $Z\to\nu\bar{\nu}$ decay, and $\Delta\phi_{i,j}$ the azimuthal separations between the corresponding objects.

\begin{figure}[htbp]
\centering
\includegraphics[width=0.9\columnwidth]{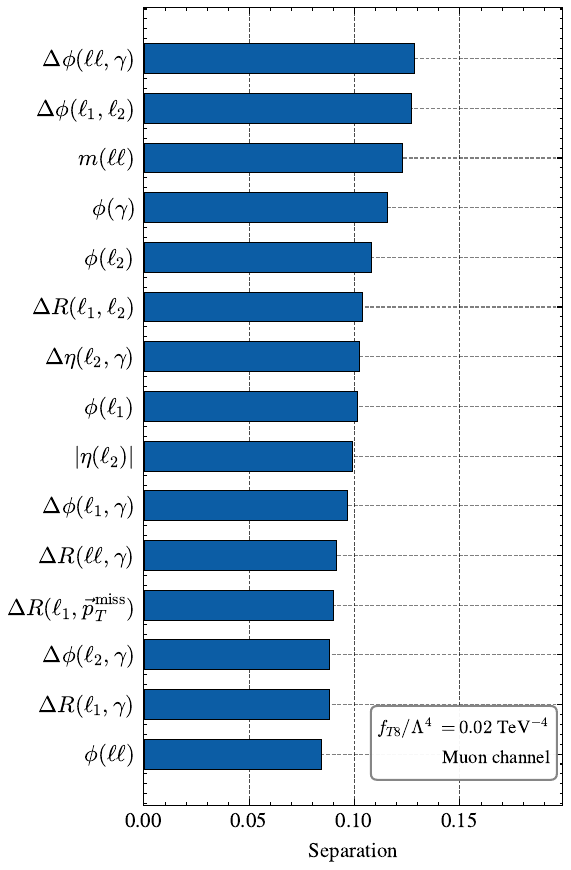}
\caption{Ranking of the input variables according to their separation power for the $f_{T8}/\Lambda^{4}=0.02$~TeV$^{-4}$ training in the muon channel. The 15 highest-ranked of the 41 input variables, ordered by the TMVA method-unspecific separation ranking, are shown.}
\label{fig:varimp}
\end{figure}

In addition to the kinematics of the reconstructed objects, three composite variables are constructed: the scalar sum of the transverse momenta $S_T = p_T^{\ell_1}+p_T^{\ell_2}+p_T^{\gamma}+E_T^{miss}$, the transverse momentum of the $ZZ\gamma$ system $p_T^{ZZ\gamma}$ obtained from the vector sum of the leptonic $Z$ and the photon transverse momenta, and the centrality, defined as the average pseudorapidity of the final-state objects. The multivariate models are trained on 41 input variables in total, comprising the transverse momenta, pseudorapidities and azimuthal angles of the selected leptons and photon, the kinematics of the reconstructed leptonic $Z$ boson, $m_{\ell\ell}$, the angular separations between the final-state objects, and the composite variables defined above; the complete list is given in Table~\ref{tab:mva_variables}. Figure~\ref{fig:varimp} shows the corresponding ranking for the $f_{T8}/\Lambda^{4}=0.02$~TeV$^{-4}$ training in the muon channel.

\begin{tableorg}[htbp]
\centering
\caption{Input variables used in the multivariate analysis, grouped by physics
object. The leading and subleading leptons of the same-flavor pair are denoted
$\ell_1$ and $\ell_2$, the reconstructed leptonic $Z$ boson $Z_{\ell\ell}$, the
photon $\gamma$, and the missing transverse momentum $E_T^{miss}$. In total 41
variables are used; $\Delta R$, $\Delta\eta$ and $\Delta\phi$ denote the angular
separations between the corresponding objects. The total transverse mass
$M_T^{tot}$ [Eq.~(\ref{eq:mttot})] is used as a spectator only.}
\label{tab:mva_variables}
\begin{adjustbox}{max width=\columnwidth}
\begin{tabular}{l l}
\toprule
Group & Variables \\
\midrule
Missing energy & $E_T^{miss}$, $\phi(E_T^{miss})$ \\
Leptons & $p_T$, $|\eta|$, $\phi$ of $\ell_1$, $\ell_2$ and the $\ell\ell$ system \\
Photon & $p_T^{\gamma}$, $|\eta^{\gamma}|$, $\phi^{\gamma}$ \\
$Z_{\ell\ell}$ boson & $m_{\ell\ell}$, $p_T$, $|\eta|$, $\phi$ \\
Angular separations & $\Delta R,\ \Delta\eta,\ \Delta\phi$ of \\
 & $(\ell_1\ell_2)$, $(\ell_1\gamma)$, $(\ell_2\gamma)$, $(Z_{\ell\ell}\gamma)$, \\
 & and $\Delta R,\ \Delta\phi$ of each object with $E_T^{miss}$ \\
Composite & $S_T$, $p_T^{ZZ\gamma}$, centrality \\
Spectator & $M_T^{tot}$ \\
\bottomrule
\end{tabular}
\end{adjustbox}
\end{tableorg}

An operator-dependent upper bound on $M_T^{tot}$ is then required to preserve unitarity. For each simulated coupling point, the energy at which the zeroth partial wave of on-shell $VV \rightarrow VV$ scattering saturates the unitarity condition $|\mathrm{Re}\,a_{0}|=0.5$ is evaluated with the VBFNLO form factor utility \texttt{calc\_formfactor~v1.4.0}~\cite{Arnold:2008rz,Baglio:2014uba}, and this energy is taken directly as the bound on $M_T^{tot}$ without applying the associated form factor. Events with $M_T^{tot}$ above this bound are removed from both the training and the evaluation, and $M_T^{tot}$ itself is not used as a training input; it is retained as a spectator variable from which the final limits are extracted. The resulting bounds are given in Table~\ref{tab:unitarity}.

\begin{table}[htbp]
\centering
\caption{Operator-dependent unitarity bound on $M_T^{tot}$ applied in the analysis. For each operator the bound is evaluated per simulated coupling point from the partial-wave unitarity condition~\cite{Rauch:2016pai,Covarelli:2021gyz} and decreases monotonically as the coupling increases. The values at the smallest and the largest scanned coupling are given. The events with $M_T^{tot}$ above the bound are removed from both the training and the evaluation.}
\label{tab:unitarity}
\begin{tabular}{lcc}
\toprule
Operator & $f/\Lambda^{4}$ range [TeV$^{-4}$] & $M_T^{tot}$ bound [TeV] \\
\midrule
$f_{T0}/\Lambda^{4}$ & $0.03$--$0.21$ & $0.76$--$0.47$ \\
$f_{T8}/\Lambda^{4}$ & $0.02$--$0.20$ & $7.01$--$3.94$ \\
$f_{T9}/\Lambda^{4}$ & $0.03$--$1.00$ & $7.80$--$3.24$ \\
$f_{M2}/\Lambda^{4}$ & $0.10$--$3.00$ & $5.35$--$2.28$ \\
\bottomrule
\end{tabular}
\end{table}

As Table~\ref{tab:cutflow_preML} shows, the tight $m_{\ell\ell}$ window already rejects the majority of the non-resonant $t\bar{t}\gamma$ and $WW\gamma$ backgrounds; the remaining separation between signal and background is provided by the multivariate analysis.

We evaluated three classification algorithms within the  \verb"TMVA" framework~\cite{TMVA:2007ngy}: Boosted Decision Trees (BDT), BDT with decorrelated inputs (BDTD), and Deep Neural Networks (DNN). All models were trained and evaluated using an equal 50:50 train–test split, with detailed configurations summarized in Table~\ref{tab:mva_config}.
\begin{tableorg}[htbp]
\centering
\caption{Configuration parameters for the multivariate classifiers used in the analysis. All methods are trained on identical input variables using an equal 50:50 train--test split (\texttt{SplitMode=Random}).}
\label{tab:mva_config}
\begin{adjustbox}{max width=\columnwidth}
\begin{tabular}{l l}
\toprule
Method & Configuration \\
\midrule
BDT  & 400 trees, max.\ depth 3, min.\ node size 5\%, \\
     & AdaBoost ($\beta=0.5$) with bagging, \\
     & Gini index, 20 cuts per variable \\
\midrule
BDTD & as BDT, without bagging, \\
     & input transformation: PCA + Gaussianisation \\
\midrule
DNN  & dense layers $128\!\to\!64\!\to\!32\!\to\!1$ (TANH, linear output), \\
     & Gaussian input transformation, Xavier initialization, \\
     & cross-entropy loss, ADAM (learning rate $10^{-3}$), \\
     & batch size 256, max.\ 50 epochs, early stopping (10), \\
     & validation fraction 20\% \\
\bottomrule
\end{tabular}
\end{adjustbox}
\end{tableorg}

\subsection{Statistical method}
\label{sec:statmethod}

\begin{figure}[htbp]
\centering
\subfigure[]{\includegraphics[width=0.48\columnwidth]{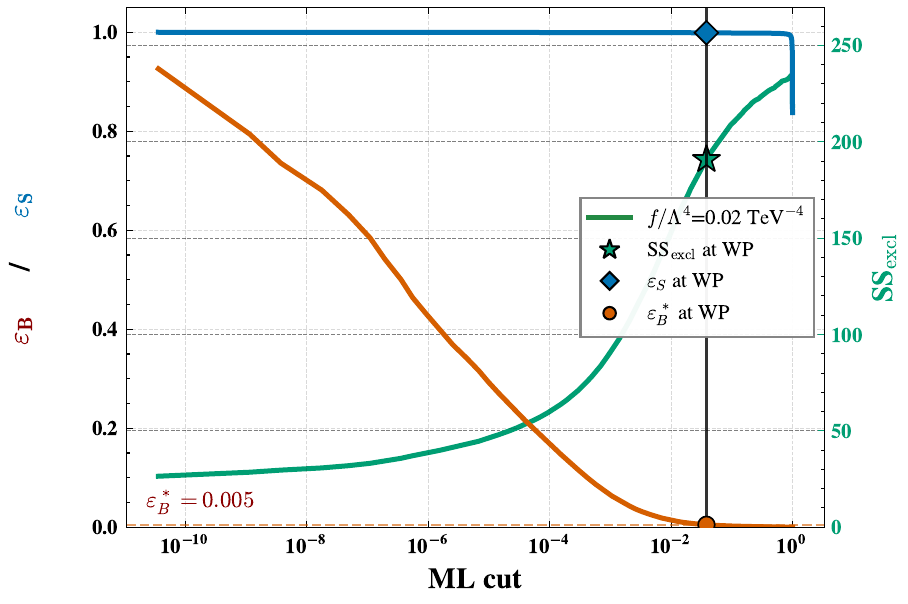}}
\subfigure[]{\includegraphics[width=0.48\columnwidth]{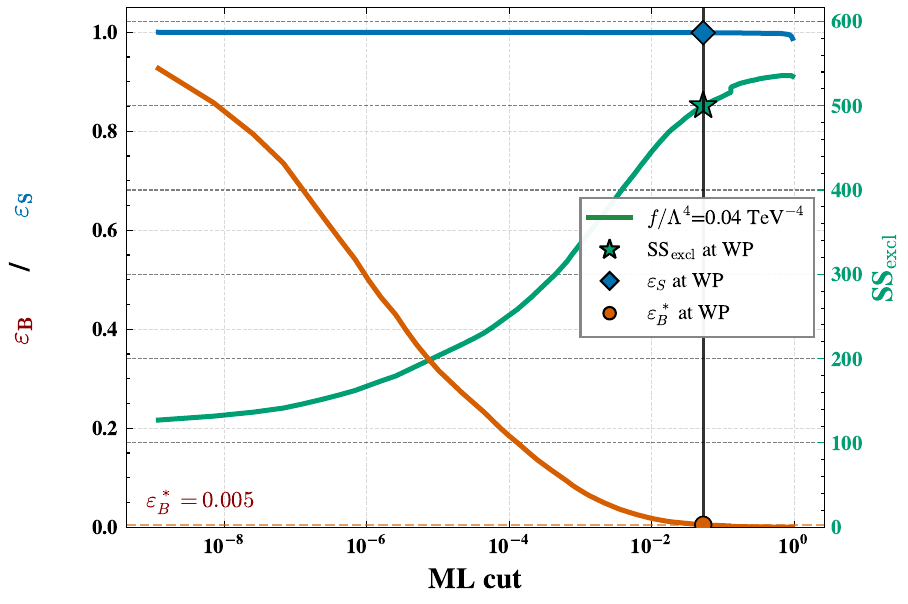}}\\[2mm]
\subfigure[]{\includegraphics[width=0.48\columnwidth]{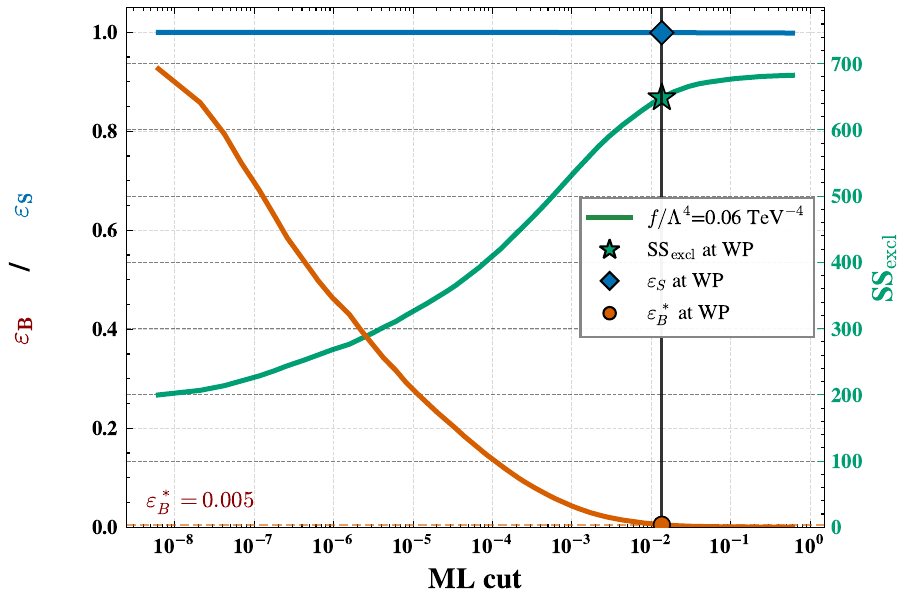}}
\subfigure[]{\includegraphics[width=0.48\columnwidth]{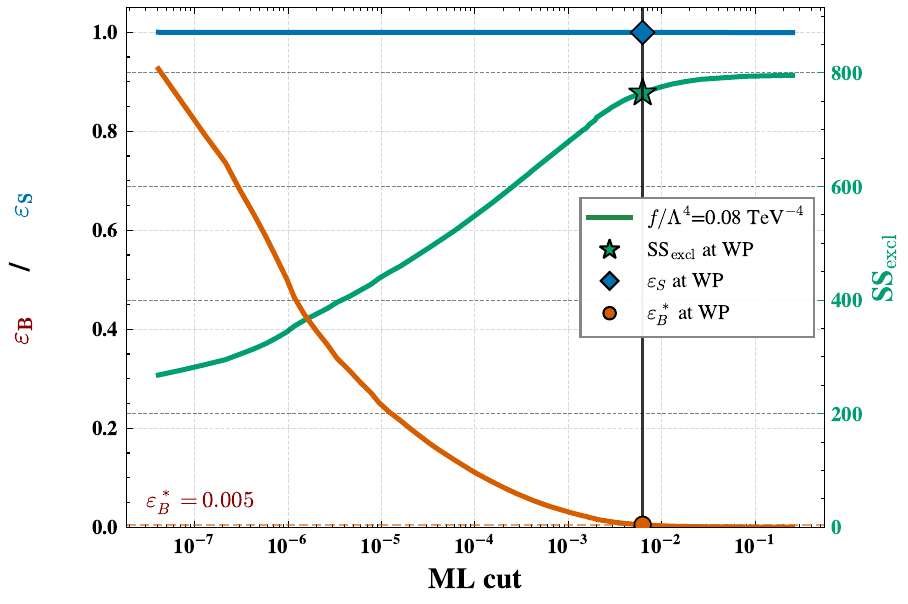}}\\[2mm]
\subfigure[]{\includegraphics[width=0.48\columnwidth]{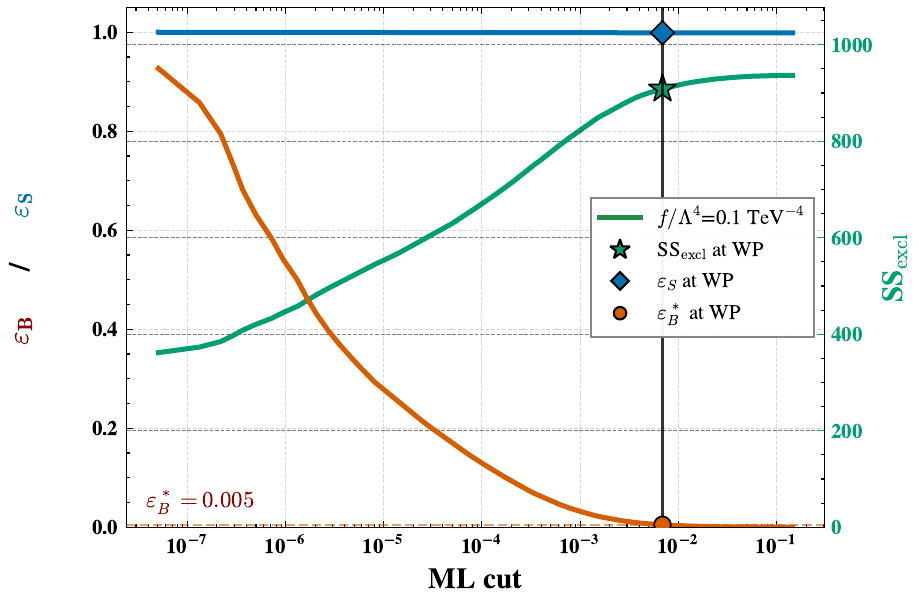}}
\subfigure[]{\includegraphics[width=0.48\columnwidth]{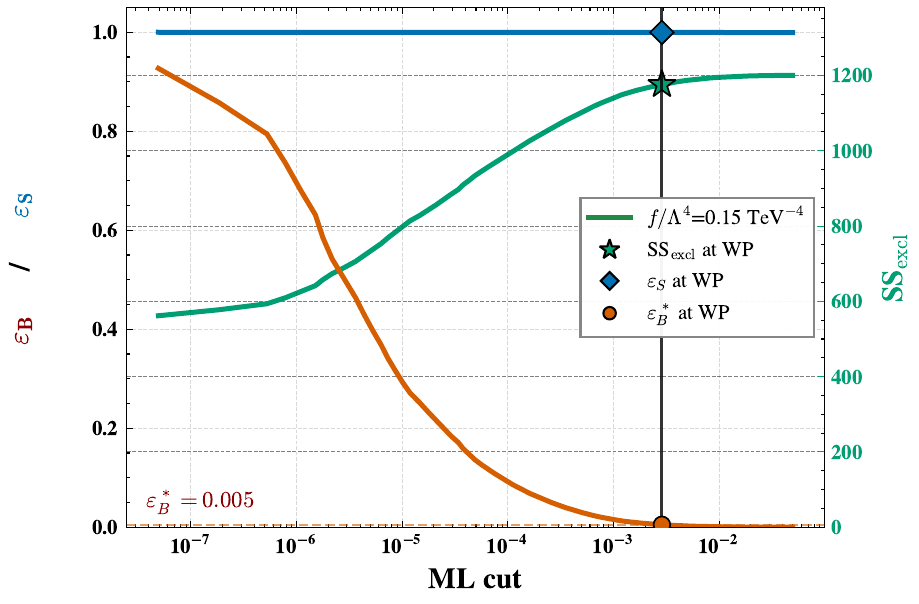}}\\[2mm]
\subfigure[]{\includegraphics[width=0.48\columnwidth]{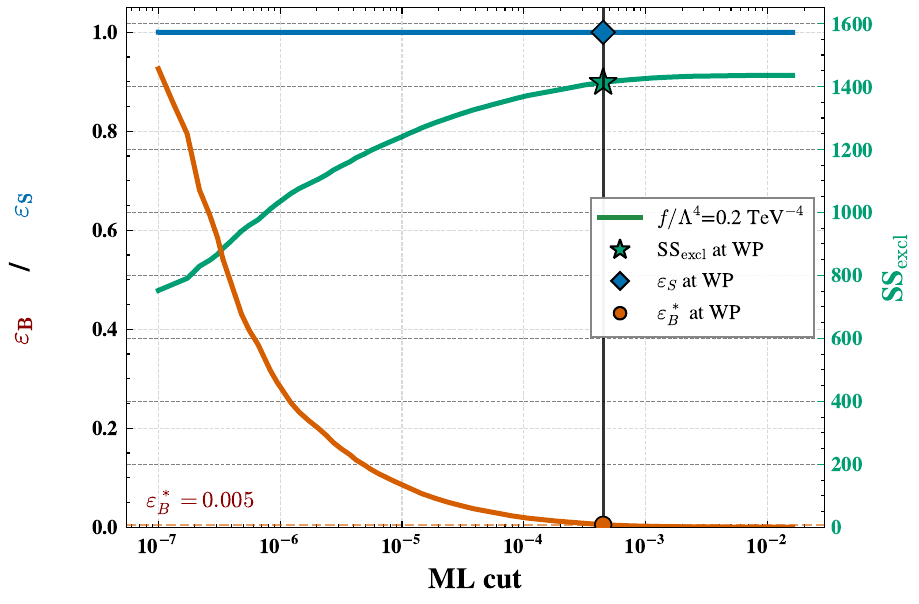}}
\caption{The working point scan for the $f_{T8}/\Lambda^{4}$ coupling with the DNN method. The background efficiency $\varepsilon_B$, the signal efficiency $\varepsilon_S$ and the exclusion significance $\text{SS}_{\text{excl}}$ are shown as a function of the ML cut, and the star indicates the selected working point corresponding to $\varepsilon_B^{\ast}=0.005$. The panels (a)--(g) correspond to the anomalous coupling values $f_{T8}/\Lambda^{4}=0.02$, 0.04, 0.06, 0.08, 0.10, 0.15 and 0.20~TeV$^{-4}$, respectively.}
\label{fig:wpscan}
\end{figure}

\begin{figure}[htbp]
\centering
\includegraphics[width=0.9\columnwidth]{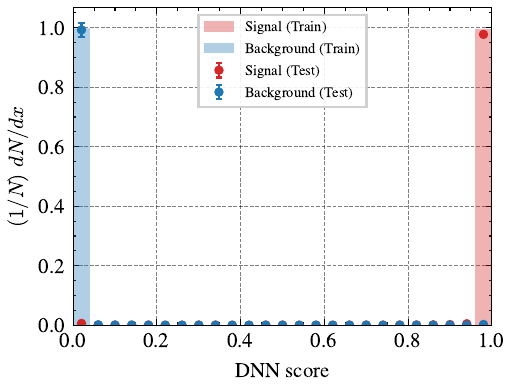}
\caption{DNN score distribution for signal and background in the $f_{T8}/\Lambda^{4}$ training, muon channel. Filled histograms show the training sample. Points with error bars show the independent testing sample. The Kolmogorov-Smirnov test gives a maximum train-test separation of $D=0.025$ (0.013) for signal (background), with $p<10^{-3}$. The working point is derived exclusively from the training sample, and the final limits are extracted from the independent testing sample.}
\label{fig:score_dist}
\end{figure}

The median expected significances are evaluated within the Asimov approximation~\cite{Cowan:2010js}. The discovery significance is:

\begin{equation}
SS_{disc} = \sqrt{2\left[(S+B)\ln\!\left(1+\frac{S}{B}\right) - S\right]},
\label{eq:ssdisc}
\end{equation}
and the exclusion significance is:

\begin{equation}
SS_{excl} = \sqrt{2\left[S - B\ln\!\left(1+\frac{S}{B}\right)\right]}.
\label{eq:ssexcl}
\end{equation}
The background systematic uncertainty ($\delta_{sys}$) is included using the equations from Ref.~\cite{Cowan:2010js}:

\begin{equation}
\begin{split}
SS_{disc} = \Bigg[2\Bigg( & (S{+}B)\ln\frac{(S{+}B)(1{+}\delta^{2}B)}{B\,(1{+}\delta^{2}(S{+}B))} \\
 & - \frac{1}{\delta^{2}}\ln\!\left(1+\frac{\delta^{2}S}{1{+}\delta^{2}B}\right)\!\Bigg)\Bigg]^{1/2}\!,
\end{split}
\label{eq:ssdisc_sys}
\end{equation}

\begin{equation}
\begin{split}
SS_{excl} = \Bigg[ 2\Bigg(S - B\ln\frac{B+S+x}{2B} \Bigg. & \\
\Bigg. {}- \frac{1}{\delta^{2}}\ln\frac{B-S+x}{2B}\Bigg) & \\
 - (B+S-x)\left(1+\frac{1}{\delta^{2}B}\right)\Bigg]^{1/2}, &
\end{split}
\label{eq:ssexcl_sys}
\end{equation}
with
$x = \sqrt{(S+B)^{2} - \frac{4\,\delta^{2}\,S\,B^{2}}{1+\delta^{2}B}},$
\noindent where $\delta = \delta_{sys}$. The 95\% confidence level (C.L.) limits are set at $SS_{excl}=1.645$~\cite{Read:2002hq}, and the $3\sigma$/$5\sigma$ reaches at $SS_{disc}=3,5$ values.
We derived the working point entirely from the training data by scanning the MVA cut to find the threshold that minimizes the median expected 95\% C.L. limit, i.e.\ the $SS_{\text{excl}}=1.645$ condition of Eq.~(\ref{eq:ssexcl}). The scan is performed jointly over all operators and all simulated coupling points rather than separately for each point. The resulting optimal background efficiency, denoted as $\varepsilon_B^{\ast}$, is nearly identical in every case; therefore, a single nominal working point of $\varepsilon_B^{\ast}=0.005$ is adopted for all results. Figure~\ref{fig:wpscan} illustrates this scan for $f_{T8}/\Lambda^{4}$. This cut is then applied to the independent testing sample to extract the final results. This procedure mitigates the impact of MVA overtraining on the extracted limits. Figure~\ref{fig:score_dist} shows the DNN score distribution for signal and background in the muon channel, with the training and testing samples overlaid. The close agreement between the two confirms that overtraining has a negligible effect on the extracted limits.

\begin{figure}[htbp]
\centering
\subfigure[$f_{T0}/\Lambda^{4}$, before DNN cut]{\includegraphics[width=0.49\columnwidth]{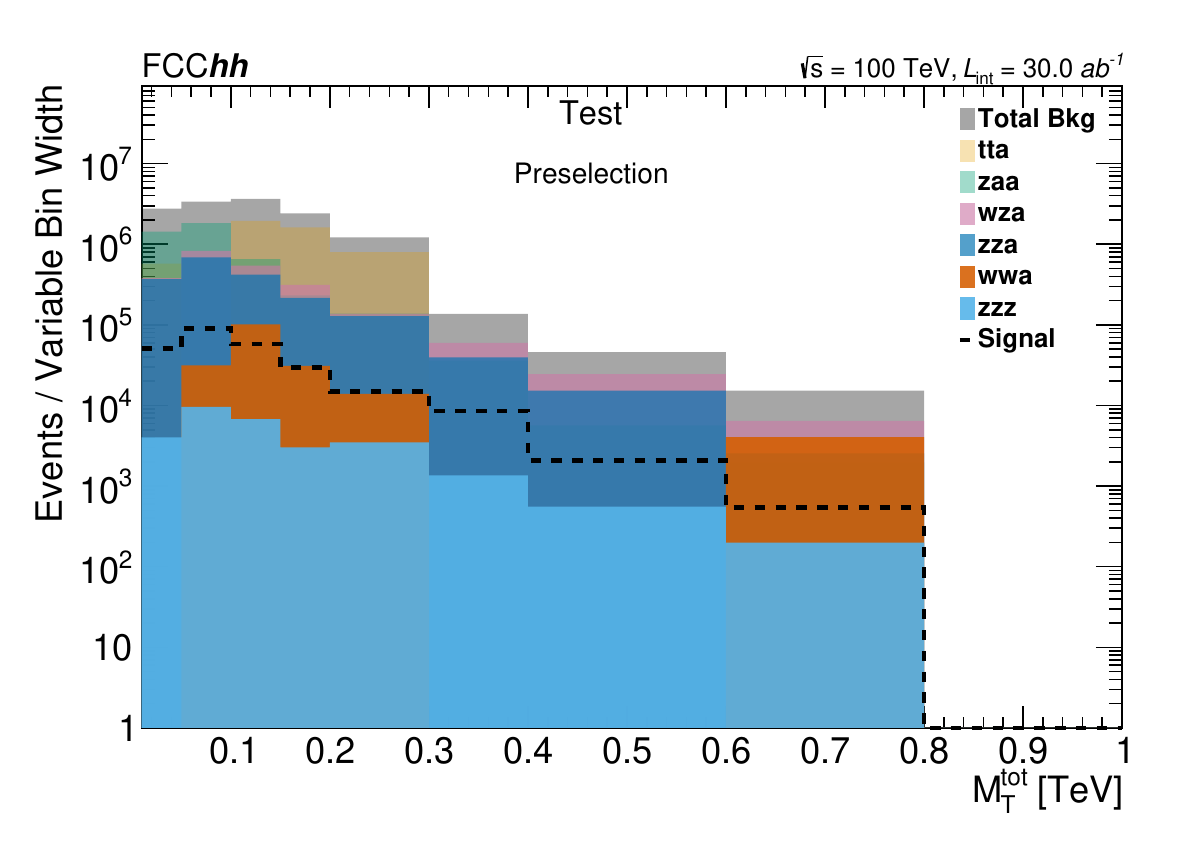}}
\subfigure[$f_{T0}/\Lambda^{4}$, after DNN cut]{\includegraphics[width=0.49\columnwidth]{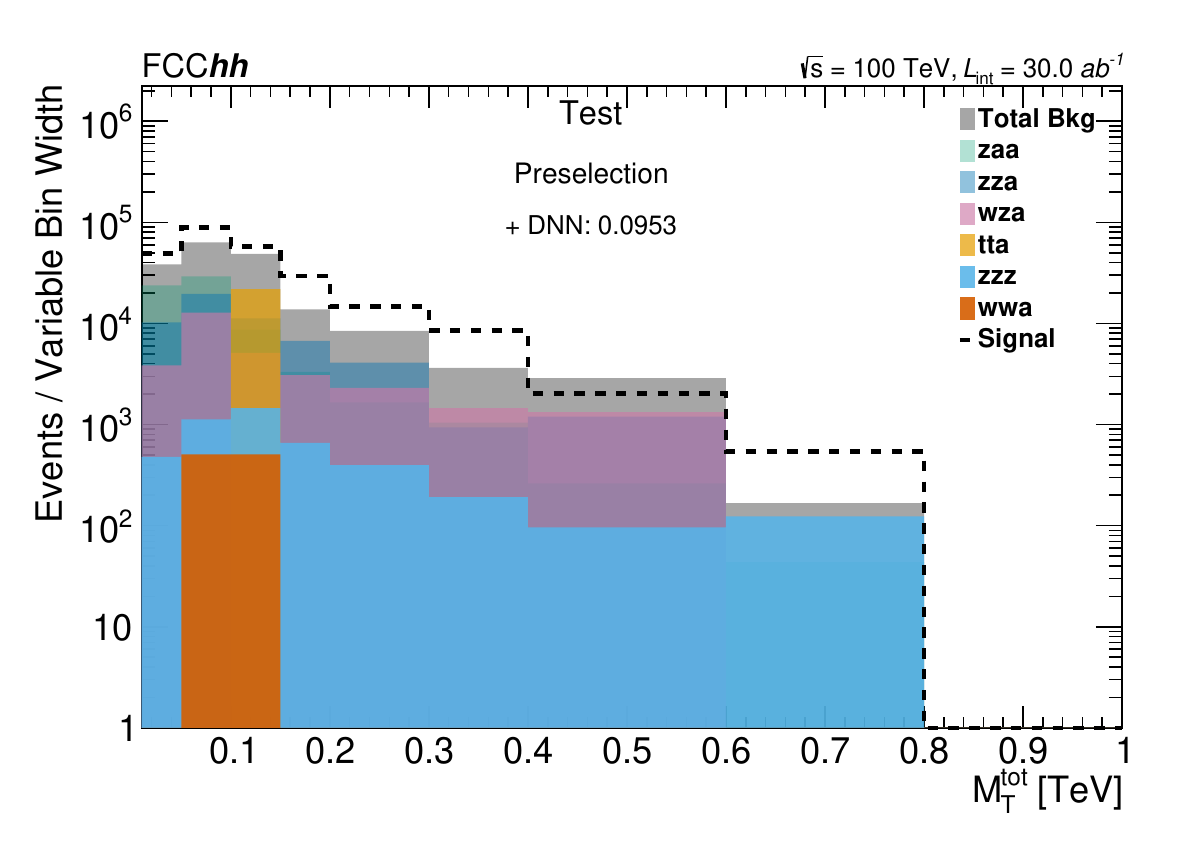}}\\[2mm]
\subfigure[$f_{T8}/\Lambda^{4}$, before DNN cut]{\includegraphics[width=0.49\columnwidth]{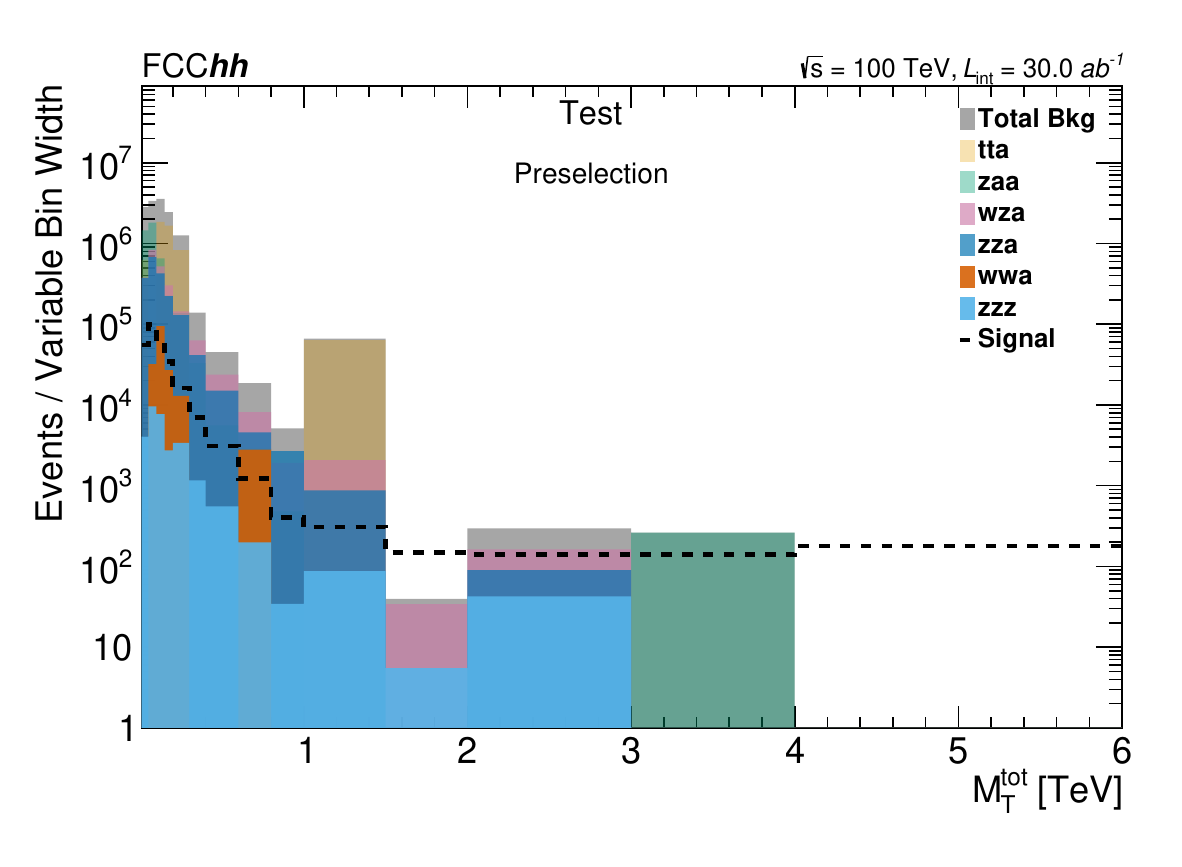}}
\subfigure[$f_{T8}/\Lambda^{4}$, after DNN cut]{\includegraphics[width=0.49\columnwidth]{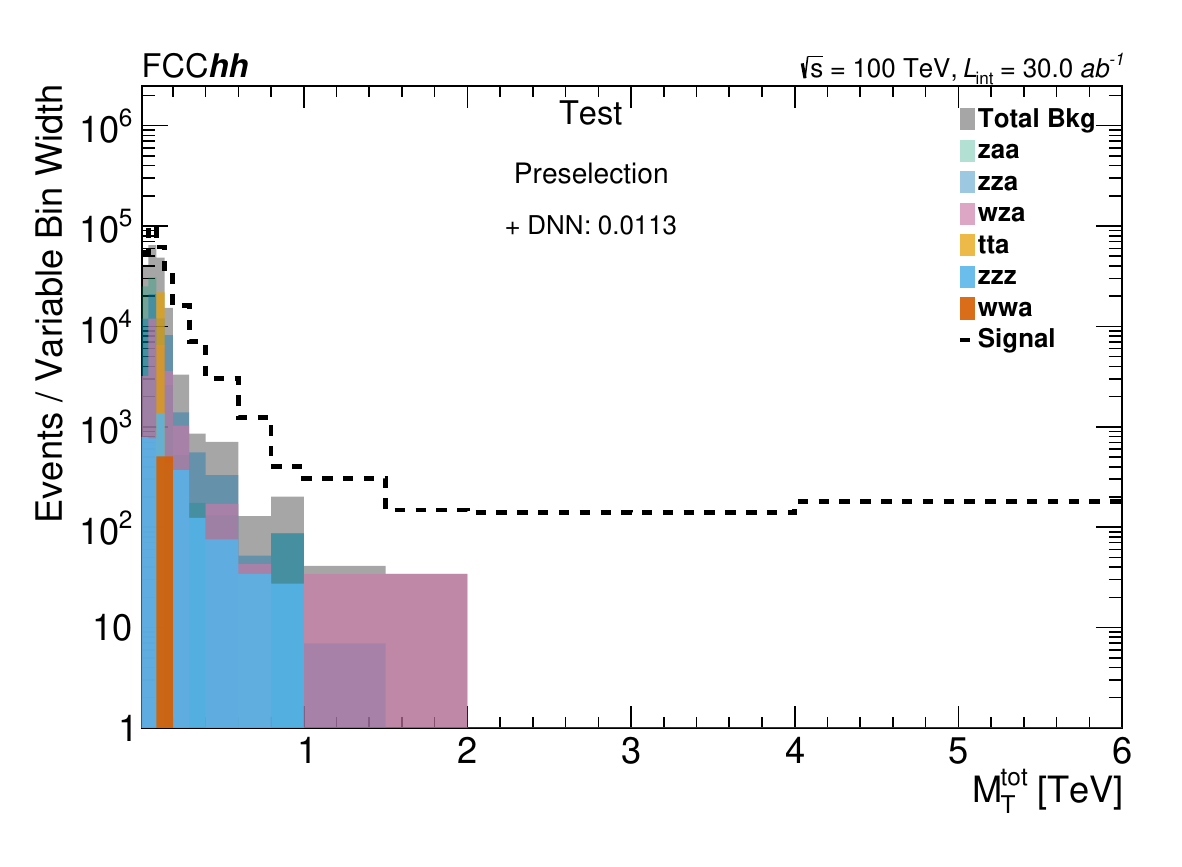}}\\[2mm]
\subfigure[$f_{T9}/\Lambda^{4}$, before DNN cut]{\includegraphics[width=0.49\columnwidth]{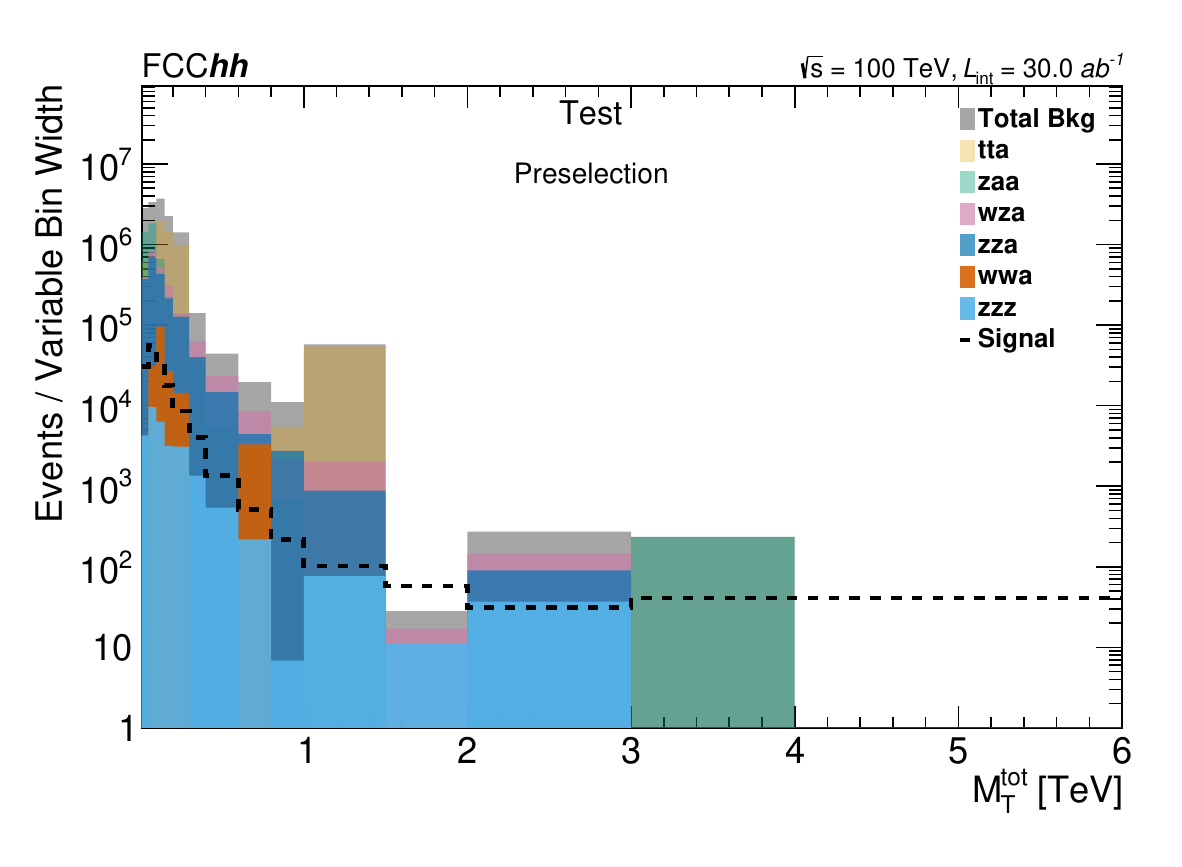}}
\subfigure[$f_{T9}/\Lambda^{4}$, after DNN cut]{\includegraphics[width=0.49\columnwidth]{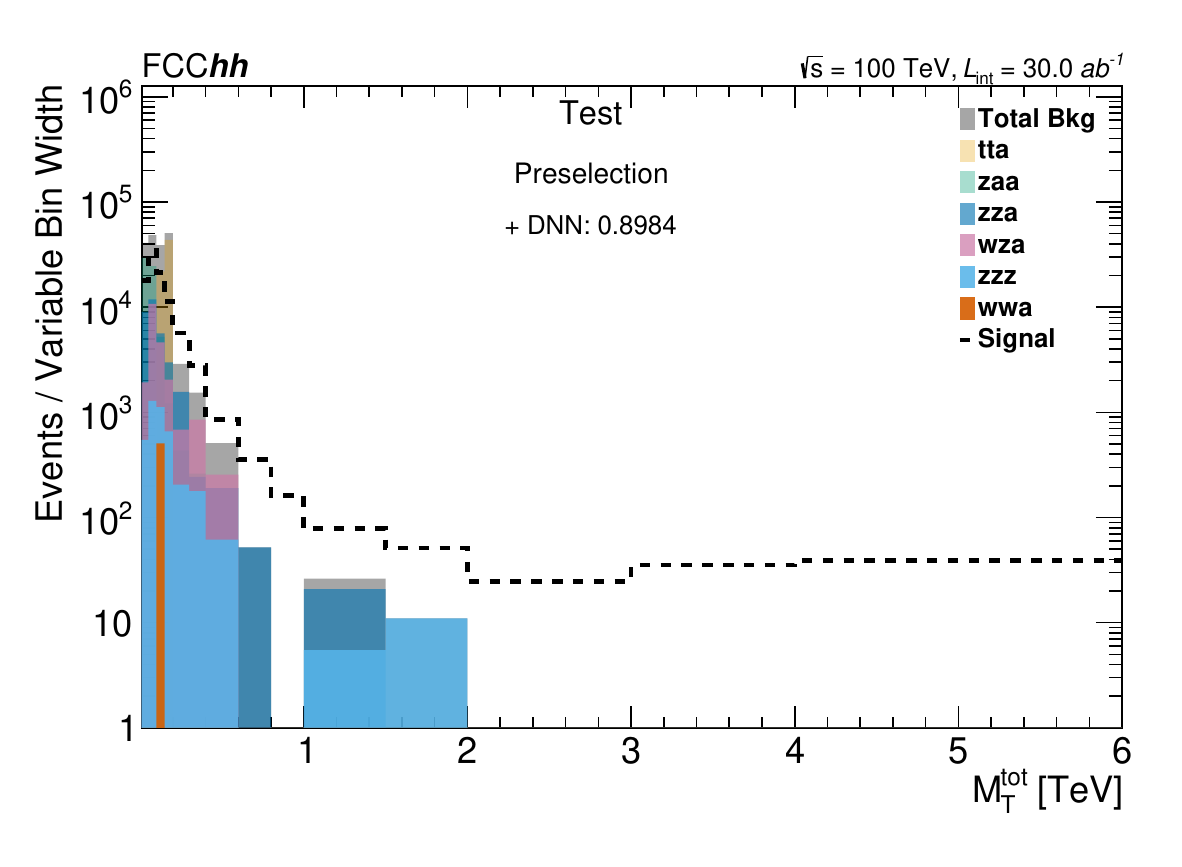}}\\[2mm]
\subfigure[$f_{M2}/\Lambda^{4}$, before DNN cut]{\includegraphics[width=0.49\columnwidth]{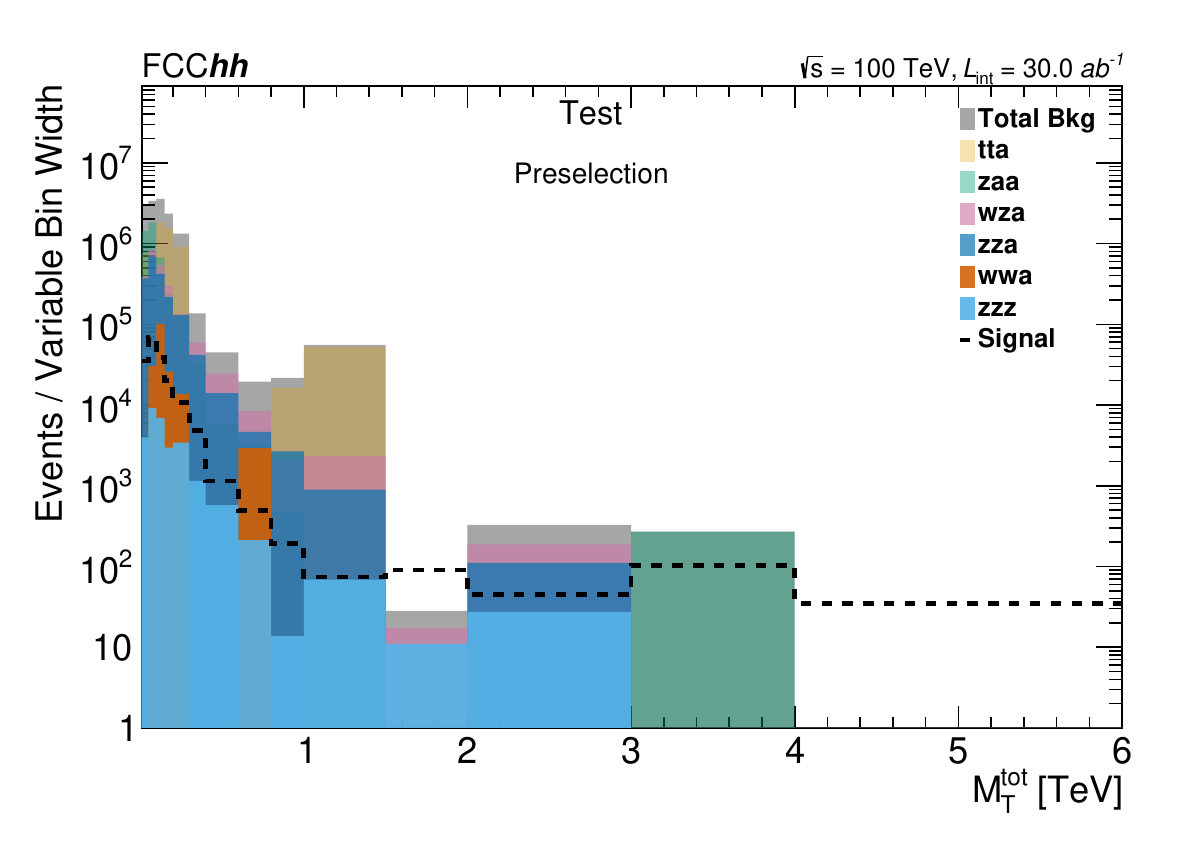}}
\subfigure[$f_{M2}/\Lambda^{4}$, after DNN cut]{\includegraphics[width=0.49\columnwidth]{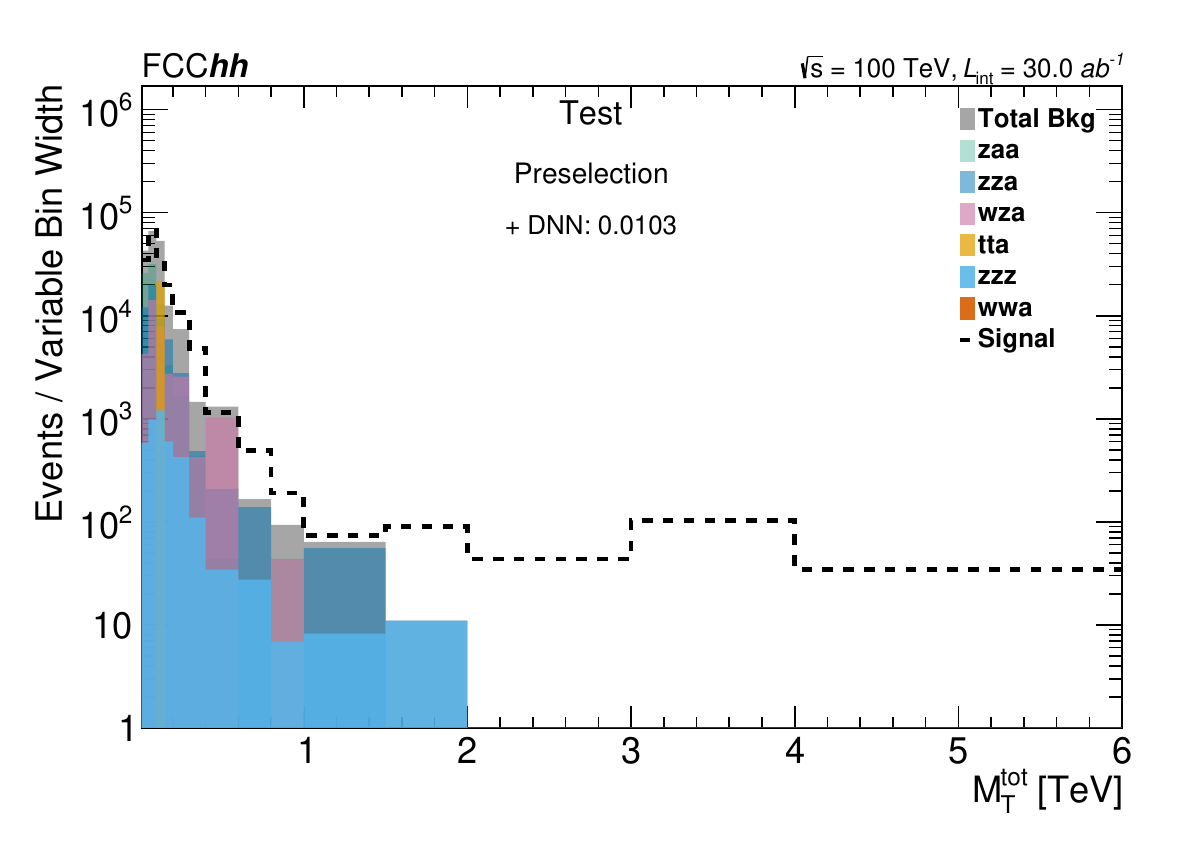}}
\caption{The $M_T^{tot}$ distributions of the signal and the SM background in the muon channel before and after the DNN score cut at $\varepsilon_B^{\ast}=0.005$. The score cut suppresses the SM background while retaining the high-$M_T^{tot}$ region, where the anomalous contribution is enhanced. Each row shows one operator before (left) and after (right) the cut, with $f_{T0}/\Lambda^{4}=0.03$~TeV$^{-4}$ in (a) and (b), $f_{T8}/\Lambda^{4}=0.02$~TeV$^{-4}$ in (c) and (d), $f_{T9}/\Lambda^{4}=0.03$~TeV$^{-4}$ in (e) and (f), and $f_{M2}/\Lambda^{4}=0.1$~TeV$^{-4}$ in (g) and (h).}
\label{fig:mttot_beforeafter}
\end{figure}

The DNN provides the highest separation power. Figure~\ref{fig:mttot_beforeafter} shows the $M_T^{tot}$ spectrum in the muon channel before and after applying the DNN score cut. The score cut suppresses the SM background while retaining the high-$M_T^{tot}$ region, where the anomalous signal is concentrated. The lowest simulated coupling that exhibits visible separation is shown.

\section{Results}
\label{sec:results}

\begin{tableorg}[htbp]
\centering
\caption{Signal ($S$) and background ($B$) event yields at the working point ($\varepsilon_B^{\ast}=0.005$ score cut and unitarity $M_T^{tot}$ bound) for the combined $e+\mu$ channel, Test sample, at $L_{\mathrm{int}}=30\,\mathrm{ab}^{-1}$. Signal yields correspond to the benchmark coupling value $f$ (in $[\mathrm{TeV}]^{-4}$) with all other couplings set to zero.}
\label{tab:yields_wp}
\begin{adjustbox}{max width=\columnwidth}
\begin{tabular}{l c rr rr rr}
\toprule
 & & \multicolumn{2}{c}{BDT} & \multicolumn{2}{c}{BDTD} & \multicolumn{2}{c}{DNN} \\
\cmidrule(lr){3-4}\cmidrule(lr){5-6}\cmidrule(lr){7-8}
Coupling & $f$ & $S$ & $B$ & $S$ & $B$ & $S$ & $B$ \\
\midrule
$f_{T0}/\Lambda^4$ & 0.03 & 4\,122 & 8\,028 & 5\,270 & 7\,399 & 23\,534 & 17\,816 \\
$f_{T8}/\Lambda^4$ & 0.02 & 6\,614 & 6\,667 & 8\,398 & 7\,266 & 29\,484 & 13\,909 \\
$f_{T9}/\Lambda^4$ & 0.03 & 1\,532 & 6\,871 & 1\,878 & 7\,156 & 10\,741 & 15\,428 \\
$f_{M2}/\Lambda^4$ & 0.1 & 3\,155 & 7\,717 & 4\,087 & 7\,562 & 16\,802 & 13\,696 \\
\bottomrule
\end{tabular}
\end{adjustbox}
\end{tableorg}

\begin{figure}[htbp]
\centering
\subfigure[]{\includegraphics[width=0.49\columnwidth]{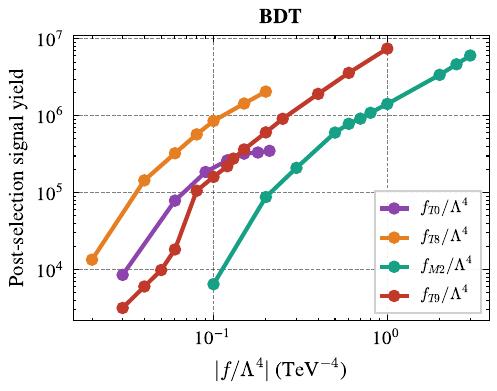}}
\subfigure[]{\includegraphics[width=0.49\columnwidth]{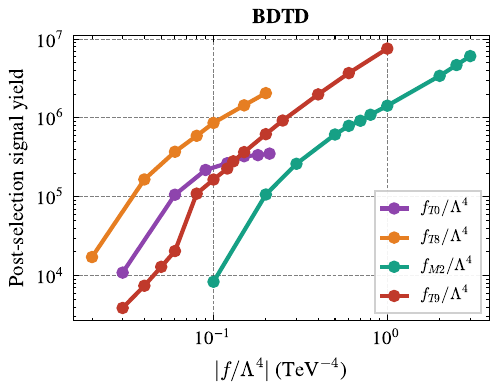}}\\[2mm]
\subfigure[]{\includegraphics[width=0.49\columnwidth]{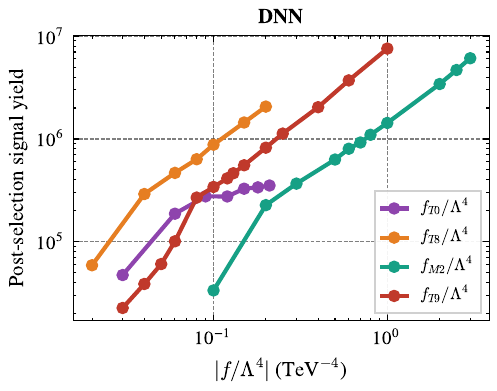}}
\caption{Post-selection signal yield as a function of the anomalous coupling $|f/\Lambda^{4}|$ (in units of TeV$^{-4}$) in the combined $e+\mu$ channel, shown on a log-log scale for the four operators $f_{T0}/\Lambda^{4}$, $f_{T8}/\Lambda^{4}$, $f_{T9}/\Lambda^{4}$ and $f_{M2}/\Lambda^{4}$. The BDT result is shown in panel (a), the BDTD result in panel (b) and the DNN result in panel (c). The near slope-two trend confirms the pure quadratic scaling used in the statistical interpretation.}
\label{fig:yieldscaling}
\end{figure}

Final limits are extracted from the independent testing sample assuming the quadratic yield model. Figure~\ref{fig:yieldscaling} shows that the post-selection yield follows the scaling across the full range of simulated coupling points, and the parameterization is extrapolated below that range to reach the final 95\% C.L. limits, which lie about an order of magnitude below the smallest simulated coupling.

\begin{figure}[htbp]
\centering
\subfigure[$f_{T0}/\Lambda^{4}$, discovery]{\includegraphics[width=0.48\columnwidth]{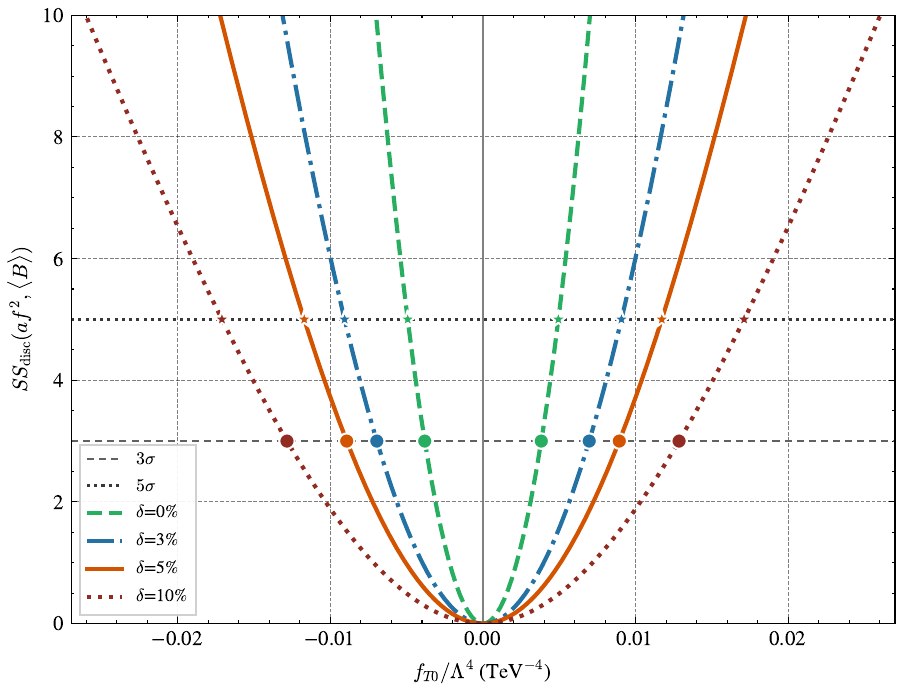}}
\subfigure[$f_{T0}/\Lambda^{4}$, exclusion]{\includegraphics[width=0.48\columnwidth]{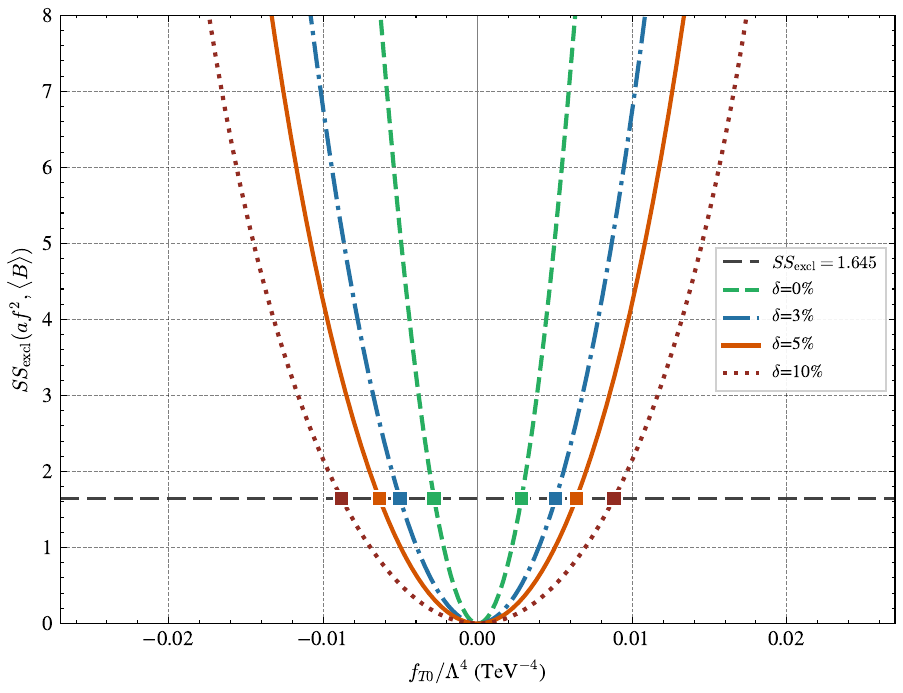}}\\[2mm]
\subfigure[$f_{T8}/\Lambda^{4}$, discovery]{\includegraphics[width=0.48\columnwidth]{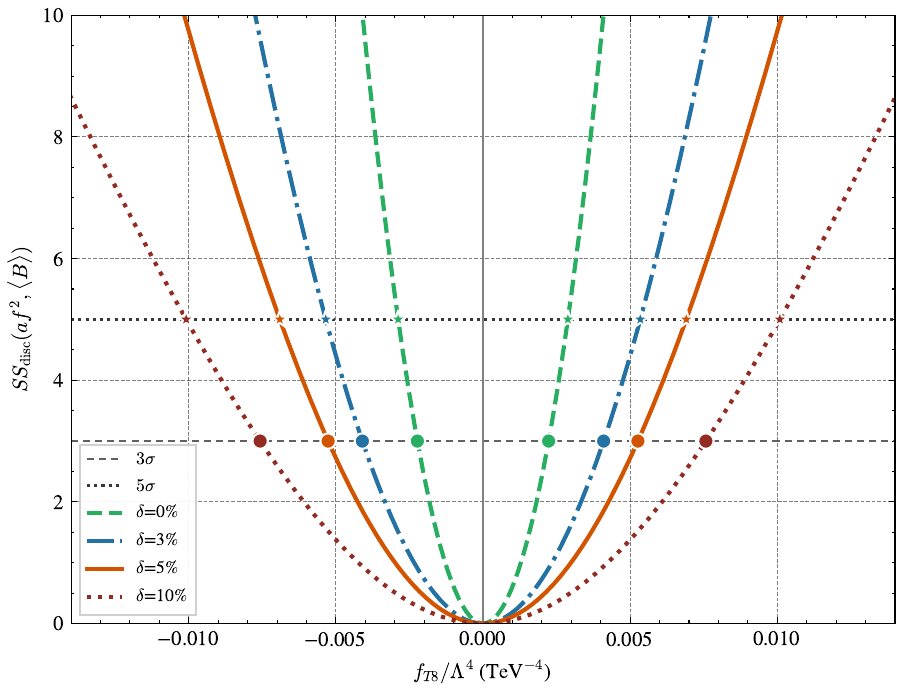}}
\subfigure[$f_{T8}/\Lambda^{4}$, exclusion]{\includegraphics[width=0.48\columnwidth]{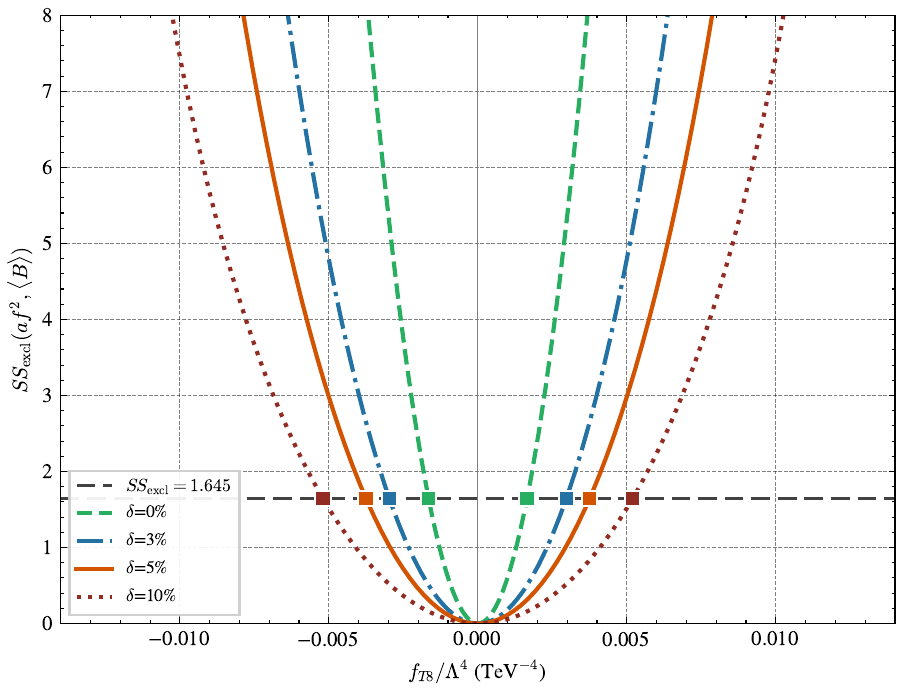}}\\[2mm]
\subfigure[$f_{T9}/\Lambda^{4}$, discovery]{\includegraphics[width=0.48\columnwidth]{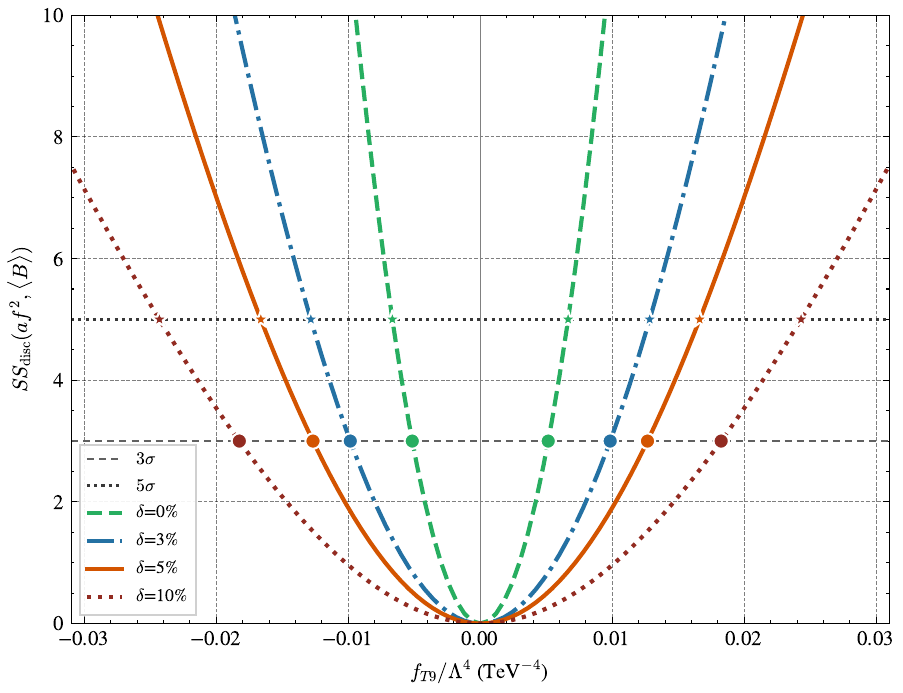}}
\subfigure[$f_{T9}/\Lambda^{4}$, exclusion]{\includegraphics[width=0.48\columnwidth]{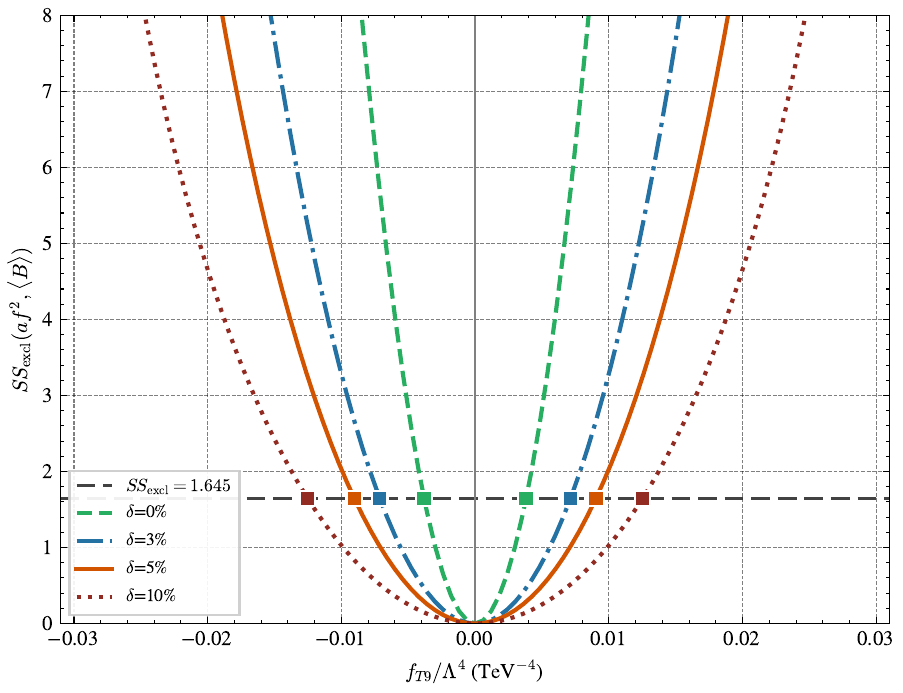}}\\[2mm]
\subfigure[$f_{M2}/\Lambda^{4}$, discovery]{\includegraphics[width=0.48\columnwidth]{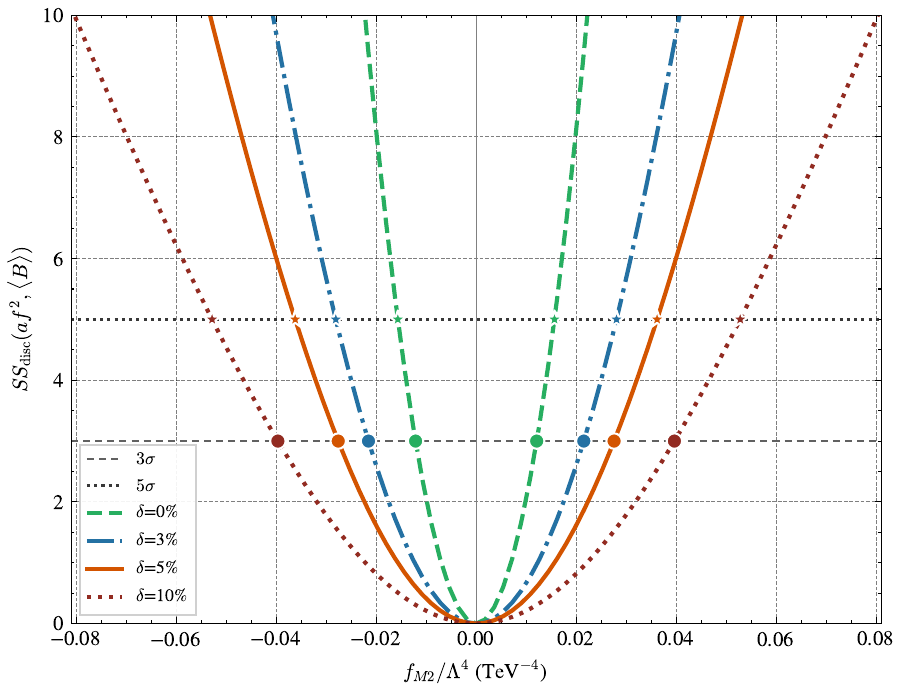}}
\subfigure[$f_{M2}/\Lambda^{4}$, exclusion]{\includegraphics[width=0.48\columnwidth]{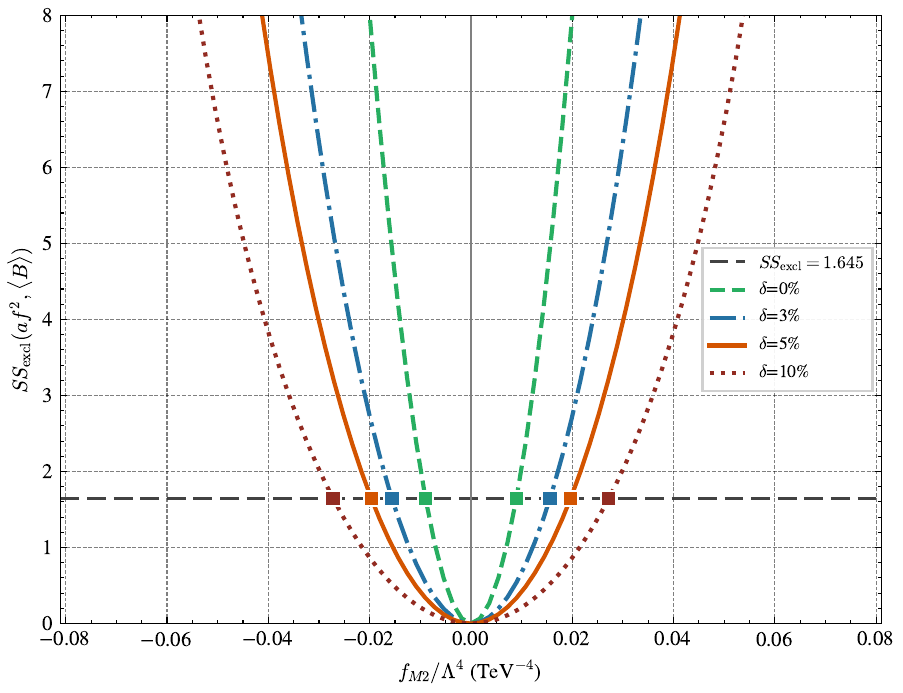}}
\caption{Median expected significance, calculated with the Asimov approximation, as a function of the anomalous coupling $f/\Lambda^{4}$ for the DNN method in the combined $e+\mu$ channel (Test sample), for background systematic uncertainties $\delta_{sys}=0,3,5,10\%$. The left column shows the discovery significance $SS_{disc}$, with the $3\sigma$ and $5\sigma$ thresholds marked by circles and stars, respectively. The right column shows the exclusion significance $SS_{excl}$, with the 95\% C.L.\ threshold ($SS_{excl}=1.645$) marked by squares. The panel pairs (a,~b), (c,~d), (e,~f) and (g,~h) correspond to $f_{T0}/\Lambda^{4}$, $f_{T8}/\Lambda^{4}$, $f_{T9}/\Lambda^{4}$ and $f_{M2}/\Lambda^{4}$, respectively. The underlying $\delta_{sys}=0\%$ values are listed in Table~\ref{tab:limits_combined_peer}.}
\label{fig:ss_multiop}
\end{figure}

Table~\ref{tab:limits_combined_peer} presents the final limits. Without systematics, the DNN 95\% C.L. limits for $f_{T0}/\Lambda^{4}$, $f_{T8}/\Lambda^{4}$, $f_{T9}/\Lambda^{4}$, and $f_{M2}/\Lambda^{4}$ are $2.83\times10^{-3}$, $1.65\times10^{-3}$, $3.81\times10^{-3}$, and $8.97\times10^{-3}$~TeV$^{-4}$, respectively. Including 5\% systematic uncertainty degrades the sensitivity by approximately a factor of two (Figure~\ref{fig:ss_multiop}).
{\sloppy
The relative performance ranking of the three methods is consistent across all operators. The DNN yields the most stringent bounds across all uncertainty scenarios, as it better models complex multivariate correlations. Among the four operators, the most stringent constraints are placed on $f_{T8}/\Lambda^{4}$.
\par}
Table~\ref{tab:limits_combined_peer} lists the final limits. Without systematics, the DNN 95\% C.L. limits are $2.83\times10^{-3}$, $1.65\times10^{-3}$, $3.81\times10^{-3}$ and $8.97\times10^{-3}$~TeV$^{-4}$ for $f_{T0}/\Lambda^{4}$, $f_{T8}/\Lambda^{4}$, $f_{T9}/\Lambda^{4}$, and $f_{M2}/\Lambda^{4}$. Including a 5\% systematic uncertainty degrades the sensitivity by approximately a factor of two (Table~\ref{tab:limits_combined_peer}, Figure~\ref{fig:ss_multiop}).

\begin{table*}[htbp]
\centering
\caption{Expected sensitivity on the aQGC couplings from the $pp\to ZZ\gamma$ process ($\ell\ell\nu\bar{\nu}\gamma$ final state) combined $e+\mu$ channel, Test sample. Since the excess yield is quadratic in the coupling and therefore sign-blind, entries are given as the symmetric two-sided range $[-f,f]$, in units of $10^{-3}\,[\mathrm{TeV}]^{-4}$, obtained from $3\sigma$ and $5\sigma$ thresholds on $SS_{\mathrm{disc}}$ and the 95\% C.L.\ threshold $SS_{\mathrm{excl}}=1.645$, for systematic uncertainties $\delta_{\mathrm{sys}}=0,3,5,10\%$.}
\label{tab:limits_combined_peer}
\resizebox{\textwidth}{!}{%
\begin{tabular}{l r ccc ccc ccc}
\toprule
 & & \multicolumn{3}{c}{$3\sigma$} & \multicolumn{3}{c}{$5\sigma$} & \multicolumn{3}{c}{95\% C.L.} \\
\cmidrule(lr){3-5}\cmidrule(lr){6-8}\cmidrule(lr){9-11}
Coupling & $\delta_{\mathrm{sys}}$ & BDT & BDTD & DNN & BDT & BDTD & DNN & BDT & BDTD & DNN \\
\midrule
$f_{T0}/\Lambda^4$ & 0\% & $[-5.41,\,5.41]$ & $[-4.77,\,4.77]$ & $[-3.82,\,3.82]$ & $[-6.99,\,6.99]$ & $[-6.17,\,6.17]$ & $[-4.94,\,4.94]$ & $[-4.00,\,4.00]$ & $[-3.53,\,3.53]$ & $[-2.83,\,2.83]$ \\
 & 3\% & $[-9.13,\,9.13]$ & $[-8.01,\,8.01]$ & $[-6.96,\,6.96]$ & $[-11.9,\,11.9]$ & $[-10.4,\,10.4]$ & $[-9.07,\,9.07]$ & $[-6.62,\,6.62]$ & $[-5.80,\,5.80]$ & $[-5.04,\,5.04]$ \\
 & 5\% & $[-11.6,\,11.6]$ & $[-10.2,\,10.2]$ & $[-8.93,\,8.93]$ & $[-15.3,\,15.3]$ & $[-13.4,\,13.4]$ & $[-11.7,\,11.7]$ & $[-8.31,\,8.31]$ & $[-7.28,\,7.28]$ & $[-6.37,\,6.37]$ \\
 & 10\% & $[-16.7,\,16.7]$ & $[-14.6,\,14.6]$ & $[-12.8,\,12.8]$ & $[-22.3,\,22.3]$ & $[-19.5,\,19.5]$ & $[-17.1,\,17.1]$ & $[-11.5,\,11.5]$ & $[-10.1,\,10.1]$ & $[-8.82,\,8.82]$ \\
\midrule
$f_{T8}/\Lambda^4$ & 0\% & $[-2.68,\,2.68]$ & $[-2.70,\,2.70]$ & $[-2.23,\,2.23]$ & $[-3.47,\,3.47]$ & $[-3.49,\,3.49]$ & $[-2.88,\,2.88]$ & $[-1.99,\,1.99]$ & $[-2.00,\,2.00]$ & $[-1.65,\,1.65]$ \\
 & 3\% & $[-4.48,\,4.48]$ & $[-4.58,\,4.58]$ & $[-4.10,\,4.10]$ & $[-5.84,\,5.84]$ & $[-5.97,\,5.97]$ & $[-5.34,\,5.34]$ & $[-3.25,\,3.25]$ & $[-3.32,\,3.32]$ & $[-2.97,\,2.97]$ \\
 & 5\% & $[-5.71,\,5.71]$ & $[-5.84,\,5.84]$ & $[-5.26,\,5.26]$ & $[-7.49,\,7.49]$ & $[-7.67,\,7.67]$ & $[-6.91,\,6.91]$ & $[-4.07,\,4.07]$ & $[-4.17,\,4.17]$ & $[-3.75,\,3.75]$ \\
 & 10\% & $[-8.19,\,8.19]$ & $[-8.38,\,8.38]$ & $[-7.57,\,7.57]$ & $[-10.9,\,10.9]$ & $[-11.2,\,11.2]$ & $[-10.1,\,10.1]$ & $[-5.62,\,5.62]$ & $[-5.75,\,5.75]$ & $[-5.20,\,5.20]$ \\
\midrule
$f_{T9}/\Lambda^4$ & 0\% & $[-6.06,\,6.06]$ & $[-6.11,\,6.11]$ & $[-5.15,\,5.15]$ & $[-7.84,\,7.84]$ & $[-7.90,\,7.90]$ & $[-6.66,\,6.66]$ & $[-4.49,\,4.49]$ & $[-4.52,\,4.52]$ & $[-3.81,\,3.81]$ \\
 & 3\% & $[-10.1,\,10.1]$ & $[-10.3,\,10.3]$ & $[-9.84,\,9.84]$ & $[-13.1,\,13.1]$ & $[-13.5,\,13.5]$ & $[-12.8,\,12.8]$ & $[-7.30,\,7.30]$ & $[-7.48,\,7.48]$ & $[-7.13,\,7.13]$ \\
 & 5\% & $[-12.8,\,12.8]$ & $[-13.2,\,13.2]$ & $[-12.7,\,12.7]$ & $[-16.8,\,16.8]$ & $[-17.3,\,17.3]$ & $[-16.6,\,16.6]$ & $[-9.16,\,9.16]$ & $[-9.39,\,9.39]$ & $[-9.03,\,9.03]$ \\
 & 10\% & $[-18.4,\,18.4]$ & $[-18.9,\,18.9]$ & $[-18.2,\,18.2]$ & $[-24.5,\,24.5]$ & $[-25.2,\,25.2]$ & $[-24.3,\,24.3]$ & $[-12.6,\,12.6]$ & $[-13.0,\,13.0]$ & $[-12.5,\,12.5]$ \\
\midrule
$f_{M2}/\Lambda^4$ & 0\% & $[-16.4,\,16.4]$ & $[-16.4,\,16.4]$ & $[-12.1,\,12.1]$ & $[-21.2,\,21.2]$ & $[-21.2,\,21.2]$ & $[-15.7,\,15.7]$ & $[-12.2,\,12.2]$ & $[-12.1,\,12.1]$ & $[-8.97,\,8.97]$ \\
 & 3\% & $[-27.4,\,27.4]$ & $[-27.6,\,27.6]$ & $[-21.5,\,21.5]$ & $[-35.7,\,35.7]$ & $[-36.0,\,36.0]$ & $[-28.1,\,28.1]$ & $[-19.9,\,19.9]$ & $[-20.0,\,20.0]$ & $[-15.6,\,15.6]$ \\
 & 5\% & $[-34.9,\,34.9]$ & $[-35.3,\,35.3]$ & $[-27.6,\,27.6]$ & $[-45.8,\,45.8]$ & $[-46.3,\,46.3]$ & $[-36.2,\,36.2]$ & $[-24.9,\,24.9]$ & $[-25.2,\,25.2]$ & $[-19.7,\,19.7]$ \\
 & 10\% & $[-50.1,\,50.1]$ & $[-50.6,\,50.6]$ & $[-39.6,\,39.6]$ & $[-66.7,\,66.7]$ & $[-67.4,\,67.4]$ & $[-52.8,\,52.8]$ & $[-34.4,\,34.4]$ & $[-34.7,\,34.7]$ & $[-27.2,\,27.2]$ \\
\bottomrule
\end{tabular}%
} 
\end{table*}

Nevertheless, the FCC-hh projections improve upon current LHC bounds by more than an order of magnitude.
The muon channel yields better limits than the electron channel due to higher preselection efficiency as summarized in Table~\ref{tab:channel_dilution}. The combined $e+\mu$ channel improves the final limits by approximately $10\mbox{-}14\%$.
\begin{tableorg}[htbp]
\centering
\caption{The obtained 95\% C.L. limits, given as the symmetric two-sided range $[-f,f]$ (in units of $10^{-3}\,[\mathrm{TeV}]^{-4}$)
with the DNN method at $\delta_{sys}=0$ for the electron, muon and combined
$e+\mu$ channels, obtained from the testing sample.}
\label{tab:channel_dilution}
\begin{adjustbox}{max width=\columnwidth}
\begin{tabular}{l c c c}
\toprule
Coupling & $e$ & $\mu$ & $e+\mu$ \\
\midrule
$f_{T0}/\Lambda^4$ & $[-3.69,\,3.69]$ & $[-3.13,\,3.13]$ & \textbf{$[-2.83,\,2.83]$} \\
$f_{T8}/\Lambda^4$ & $[-2.12,\,2.12]$ & $[-1.86,\,1.86]$ & \textbf{$[-1.65,\,1.65]$} \\
$f_{T9}/\Lambda^4$ & $[-4.71,\,4.71]$ & $[-4.42,\,4.42]$ & \textbf{$[-3.81,\,3.81]$} \\
$f_{M2}/\Lambda^4$ & $[-11.1,\,11.1]$ & $[-10.3,\,10.3]$ & \textbf{$[-8.97,\,8.97]$} \\
\bottomrule
\end{tabular}
\end{adjustbox}
\end{tableorg}

As shown in Table~\ref{tab:lhc_comparison}, the FCC-hh provides a significant improvement over the current LHC limits. Without systematic uncertainties, it tightens the bounds by factors of about 33, 36, 34, and 78 for $f_{T0}/\Lambda^{4}$, $f_{T8}/\Lambda^{4}$, $f_{T9}/\Lambda^{4}$, and $f_{M2}/\Lambda^{4}$, respectively. Even with a 5\% systematic uncertainty, the improvement factors remain approximately 14, 16, 14, and 36. Because the current LHC limits are derived from various physical processes, this comparison is only an overall indicator of the 100~TeV machine's sensitivity reach. The two sets of bounds are also obtained under different unitarity conditions. The $f_{T8}/\Lambda^4$ and $f_{T9}/\Lambda^4$ limits shown are those reported in the region where unitarity is not preserved, whereas in the region where it is preserved the same analysis reports $520$ and $790$ in the same units, at cut-off scales of $1.7$ and $1.9$~TeV~\cite{ATLAS:2022nru}. For $f_{T0}/\Lambda^4$ and $f_{M2}/\Lambda^4$, Ref.~\cite{CMS:2025dbm} reports that applying a clipping unitarization scheme has only a small impact on the resulting bounds. The limits obtained in this work are extracted with partial-wave unitarity enforced through the operator-dependent $M_T^{tot}$ bound of Table~\ref{tab:unitarity}.

\begin{tableorg}[htbp]
\centering
\caption{Comparison of the obtained 95\% C.L. limits (DNN, combined $e+\mu$
channel, in units of $10^{-3}\,[\mathrm{TeV}]^{-4}$), given as the symmetric two-sided range $[-f,f]$, with the most
stringent current LHC limits, for $\delta_{\mathrm{sys}}=0\%$ and $5\%$. The LHC bounds on $f_{T8}/\Lambda^4$ and
$f_{T9}/\Lambda^4$ are obtained from the electroweak $Z(\nu\bar{\nu})\gamma jj$
measurement and those on $f_{T0}/\Lambda^4$ and $f_{M2}/\Lambda^4$ from vector
boson scattering in the semileptonic final states, both based on the full Run~2
dataset of $138$--$139$~fb$^{-1}$.}
\label{tab:lhc_comparison}
\begin{adjustbox}{max width=\columnwidth}
\begin{tabular}{l cc r cc}
\toprule
 & \multicolumn{2}{c}{Obtained limit} & & \multicolumn{2}{c}{Improvement} \\
\cmidrule(lr){2-3}\cmidrule(lr){5-6}
Coupling & $\delta_{\mathrm{sys}}=0\%$ & $\delta_{\mathrm{sys}}=5\%$ & Best LHC limit & $\delta_{\mathrm{sys}}=0\%$ & $\delta_{\mathrm{sys}}=5\%$ \\
\midrule
$f_{T0}/\Lambda^4$ & $[-2.83,\,2.83]$ & $[-6.37,\,6.37]$ & $[-92.1,\,78.5]$ ~\cite{CMS:2025dbm} & $\sim\!33$ & $\sim\!14$ \\
$f_{T8}/\Lambda^4$ & $[-1.65,\,1.65]$ & $[-3.75,\,3.75]$ & $[-59,\,59]$ ~\cite{ATLAS:2022nru} & $\sim\!36$ & $\sim\!16$ \\
$f_{T9}/\Lambda^4$ & $[-3.81,\,3.81]$ & $[-9.03,\,9.03]$ & $[-130,\,130]$ ~\cite{ATLAS:2022nru} & $\sim\!34$ & $\sim\!14$ \\
$f_{M2}/\Lambda^4$ & $[-8.97,\,8.97]$ & $[-19.7,\,19.7]$ & $[-703,\,703]$ ~\cite{CMS:2025dbm} & $\sim\!78$ & $\sim\!36$ \\
\bottomrule
\end{tabular}
\end{adjustbox}
\end{tableorg}

\section{Conclusion}
\label{sec:conclusion}

{\sloppy
The projected sensitivity to anomalous quartic gauge couplings in the $pp \rightarrow ZZ\gamma$ process at the FCC-hh (100~TeV, 30~ab$^{-1}$) has been investigated. The BDT, BDTD, and DNN techniques were compared with partial-wave unitarity enforced via an operator-dependent $M_T^{tot}$ bound, and the DNN was found to provide the highest separation power. Without systematic uncertainties, it yields absolute 95\% C.L. limits $|f|$ of $2.83\times 10^{-3}$, $1.65\times 10^{-3}$, $3.81\times 10^{-3}$, and $8.97\times 10^{-3}$~TeV$^{-4}$ for $f_{T0}/\Lambda^4$, $f_{T8}/\Lambda^4$, $f_{T9}/\Lambda^4$, and $f_{M2}/\Lambda^4$, respectively, tightening current LHC limits by factors of 33 to 78. Including a 5\% systematic uncertainty, the limits become $6.37\times 10^{-3}$, $3.75\times 10^{-3}$, $9.03\times 10^{-3}$, and $19.7\times 10^{-3}$~TeV$^{-4}$, still improving upon the LHC bounds by factors of 14 to 36.
\par} MVA overtraining effects are mitigated by deriving the working point exclusively from training data and extracting final limits from an independent testing sample. The limits are obtained with partial-wave unitarity enforced through an operator-dependent $M_T^{tot}$ bound.
Given recent LHC observations of triboson production involving a photon, the FCC-hh is expected to extend this sensitivity well beyond current LHC reach. Future studies may improve these constraints by incorporating additional $ZZ\gamma$ decay channels and analyzing the remaining dim-8 operators.

\backmatter

\bmhead{Acknowledgements}

We thank the Turkish Energy, Nuclear and Mineral Research Agency (TENMAK) for their support under Grant No. 2025TENMAK(CERN)A5.H3.F2-05.

\section*{Declarations}


\noindent\textbf{Conflict of interest} The authors declare that they have no conflict of interest.

\noindent\textbf{Data availability} The simulated event samples and the derived numerical results underlying this study are available from the corresponding author upon reasonable request. All software used (\texttt{MadGraph5\_aMC@NLO}, \texttt{PYTHIA}, \texttt{Delphes}, \texttt{CutLang}, \texttt{TMVA} and the VBFNLO form factor utility) is publicly available.

\bibliography{pp2zza_llvvaRef}

@article{ATLAS:2012yve,
	archiveprefix = {arXiv},
	author = {Aad, Georges and others},
	collaboration = {ATLAS},
	doi = {10.1016/j.physletb.2012.08.020},
	eprint = {1207.7214},
	journal = {Phys. Lett. B},
	pages = {1--29},
	primaryclass = {hep-ex},
	reportnumber = {CERN-PH-EP-2012-218},
	title = {{Observation of a new particle in the search for the Standard Model Higgs boson with the ATLAS detector at the LHC}},
	volume = {716},
	year = {2012}}

@article{CMS:2012qbp,
	archiveprefix = {arXiv},
	author = {Chatrchyan, Serguei and others},
	collaboration = {CMS},
	doi = {10.1016/j.physletb.2012.08.021},
	eprint = {1207.7235},
	journal = {Phys. Lett. B},
	pages = {30--61},
	primaryclass = {hep-ex},
	reportnumber = {CMS-HIG-12-028, CERN-PH-EP-2012-220},
	title = {{Observation of a New Boson at a Mass of 125 GeV with the CMS Experiment at the LHC}},
	volume = {716},
	year = {2012}}

@article{Martin:1997ns,
	archiveprefix = {arXiv},
	author = {Martin, Stephen P.},
	doi = {10.1142/9789812839657_0001},
	editor = {Kane, Gordon L.},
	eprint = {hep-ph/9709356},
	journal = {Adv. Ser. Direct. High Energy Phys.},
	pages = {1--98},
	reportnumber = {FERMILAB-PUB-97-425-T},
	title = {{A Supersymmetry primer}},
	volume = {18},
	year = {1998}}

@article{Bertone:2004pz,
	archiveprefix = {arXiv},
	author = {Bertone, Gianfranco and Hooper, Dan and Silk, Joseph},
	doi = {10.1016/j.physrep.2004.08.031},
	eprint = {hep-ph/0404175},
	journal = {Phys. Rept.},
	pages = {279--390},
	reportnumber = {FERMILAB-PUB-04-047-A},
	title = {{Particle dark matter: Evidence, candidates and constraints}},
	volume = {405},
	year = {2005}}

@article{Hagiwara:1986vm,
	author = {Hagiwara, Kaoru and Peccei, R. D. and Zeppenfeld, D. and Hikasa, K.},
	doi = {10.1016/0550-3213(87)90685-7},
	journal = {Nucl. Phys. B},
	pages = {253--307},
	reportnumber = {MAD/PH/279, DESY-86-058},
	title = {{Probing the Weak Boson Sector in e+ e- ---{\ensuremath{>}} W+ W-}},
	volume = {282},
	year = {1987}}

@article{Gounaris:1999kf,
	archiveprefix = {arXiv},
	author = {Gounaris, G. J. and Layssac, J. and Renard, F. M.},
	doi = {10.1103/PhysRevD.61.073013},
	eprint = {hep-ph/9910395},
	journal = {Phys. Rev. D},
	pages = {073013},
	reportnumber = {PM-99-39, THES-TP-99-11},
	title = {{Signatures of the anomalous $Z_{\gamma}$ and $Z Z$ production at the lepton and hadron colliders}},
	volume = {61},
	year = {2000}}

@article{Eboli:2006wa,
	archiveprefix = {arXiv},
	author = {Eboli, O. J. P. and Gonzalez-Garcia, M. C. and Mizukoshi, J. K.},
	doi = {10.1103/PhysRevD.74.073005},
	eprint = {hep-ph/0606118},
	journal = {Phys. Rev. D},
	pages = {073005},
	reportnumber = {YITP-SB-06-10, IFUSP-1620-2006},
	title = {{p p ---{\ensuremath{>}} j j e+- mu+- nu nu and j j e+- mu-+ nu nu at O( alpha(em)**6) and O(alpha(em)**4 alpha(s)**2) for the study of the quartic electroweak gauge boson vertex at CERN LHC}},
	volume = {74},
	year = {2006}}

@article{Eboli:2016kko,
	archiveprefix = {arXiv},
	author = {{\'E}boli, O. J. P. and Gonzalez-Garcia, M. C.},
	doi = {10.1103/PhysRevD.93.093013},
	eprint = {1604.03555},
	journal = {Phys. Rev. D},
	number = {9},
	pages = {093013},
	primaryclass = {hep-ph},
	reportnumber = {YITP-SB-16-09},
	title = {{Classifying the bosonic quartic couplings}},
	volume = {93},
	year = {2016}}

@article{ATLAS:2022nru,
	archiveprefix = {arXiv},
	author = {Aad, Georges and others},
	collaboration = {ATLAS},
	doi = {10.1007/JHEP06(2023)082},
	eprint = {2208.12741},
	journal = {JHEP},
	pages = {082},
	primaryclass = {hep-ex},
	reportnumber = {CERN-EP-2022-138},
	title = {{Measurement of electroweak $ Z\left(\nu \overline{\nu}\right)\gamma jj $ production and limits on anomalous quartic gauge couplings in pp collisions at $ \sqrt{s} $ = 13 TeV with the ATLAS detector}},
	volume = {06},
	year = {2023}}

@article{CMS:2025dbm,
	archiveprefix = {arXiv},
	author = {Hayrapetyan, Aram and others},
	collaboration = {CMS},
	doi = {10.1007/JHEP03(2026)022},
	eprint = {2510.00118},
	journal = {JHEP},
	pages = {022},
	primaryclass = {hep-ex},
	reportnumber = {CMS-SMP-22-011, CERN-EP-2025-147},
	title = {{Vector boson scattering and anomalous quartic couplings in final states with {\ensuremath{\ell}}{\ensuremath{\nu}}qq or {\ensuremath{\ell}}{\ensuremath{\ell}}qq plus jets using proton-proton collisions at $ \sqrt{s}=13 $ TeV}},
	volume = {03},
	year = {2026}}

@article{ATLAS:2025omi,
	archiveprefix = {arXiv},
	author = {Aad, Georges and others},
	collaboration = {ATLAS},
	doi = {10.1140/epjc/s10052-025-15217-3},
	eprint = {2503.17461},
	journal = {Eur. Phys. J. C},
	number = {4},
	pages = {433},
	primaryclass = {hep-ex},
	reportnumber = {CERN-EP-2025-050},
	title = {{Electroweak diboson production in association with a high-mass dijet system in semileptonic final states from pp collisions at $ \sqrt{s} = 13$ TeV with the ATLAS detector}},
	volume = {86},
	year = {2026}}

@article{CMS:2026vhr,
	archiveprefix = {arXiv},
	author = {Hayrapetyan, Aram and others},
	collaboration = {CMS},
	doi = {10.48550/arXiv.2604.02594},
	eprint = {2604.02594},
	month = {4},
	primaryclass = {hep-ex},
	reportnumber = {CMS-SMP-24-014, CERN-EP-2026-102},
	title = {{Evidence of ZZ$\gamma$ production and observation of $4\ell\gamma$ in proton-proton collisions at $\sqrt{s}$ = 13 TeV}},
	year = {2026}}

@article{CMS:2025oey,
	archiveprefix = {arXiv},
	author = {Chekhovsky, V. and others},
	collaboration = {CMS},
	doi = {10.1103/cm24-665b},
	eprint = {2503.21977},
	journal = {Phys. Rev. D},
	number = {1},
	pages = {012009},
	primaryclass = {hep-ex},
	reportnumber = {CMS-SMP-22-018, CERN-EP-2025-020},
	title = {{Observation of WZ$\gamma$ production and constraints on new physics scenarios in proton-proton collisions at $\sqrt{s}$ = 13 TeV}},
	volume = {112},
	year = {2025}}

@article{FCC:2018vvp,
	author = {Abada, A. and others},
	collaboration = {FCC},
	doi = {10.1140/epjst/e2019-900087-0},
	journal = {Eur. Phys. J. ST},
	number = {4},
	pages = {755--1107},
	reportnumber = {CERN-ACC-2018-0058},
	title = {{FCC-hh: The Hadron Collider}: {Future Circular Collider Conceptual Design Report Volume 3}},
	volume = {228},
	year = {2019}}

@article{FCC:2018byv,
	author = {Abada, A. and others},
	collaboration = {FCC},
	doi = {10.1140/epjc/s10052-019-6904-3},
	journal = {Eur. Phys. J. C},
	number = {6},
	pages = {474},
	reportnumber = {CERN-ACC-2018-0056},
	title = {{FCC Physics Opportunities}: {Future Circular Collider Conceptual Design Report Volume 1}},
	volume = {79},
	year = {2019}}

@article{Degrande:2012wf,
	archiveprefix = {arXiv},
	author = {Degrande, Celine and Greiner, Nicolas and Kilian, Wolfgang and Mattelaer, Olivier and Mebane, Harrison and Stelzer, Tim and Willenbrock, Scott and Zhang, Cen},
	doi = {10.1016/j.aop.2013.04.016},
	eprint = {1205.4231},
	journal = {Annals Phys.},
	pages = {21--32},
	primaryclass = {hep-ph},
	reportnumber = {MPP-2011-149, SI-HEP-2011-17, CP3-12-25},
	title = {{Effective Field Theory: A Modern Approach to Anomalous Couplings}},
	volume = {335},
	year = {2013}}

@article{Gutierrez-Rodriguez:2025wcy,
	archiveprefix = {arXiv},
	author = {Gutierrez-Rodr{\i}guez, A. and Cetinkaya, V. and Koksal, M. and Gurkanli, E. and Ari, V. and Hernandez-Ru{\i}z, M. A.},
	doi = {10.1140/epjp/s13360-025-06400-2},
	eprint = {2312.06329},
	journal = {Eur. Phys. J. Plus},
	pages = {5},
	primaryclass = {hep-ph},
	title = {{The future muon collider for the research of the anomalous neutral quartic $Z\gamma\gamma\gamma$, $ZZ\gamma\gamma$, and $ZZZ\gamma$ couplings}},
	volume = {140},
	year = {2025}}

@article{Senol:2018cks,
	archiveprefix = {arXiv},
	author = {Senol, A. and Denizli, H. and Yilmaz, A. and Turk Cakir, I. and Oyulmaz, K. Y. and Karadeniz, O. and Cakir, O.},
	doi = {10.1016/j.nuclphysb.2018.08.018},
	eprint = {1805.03475},
	journal = {Nucl. Phys. B},
	pages = {365--376},
	primaryclass = {hep-ph},
	title = {{Probing the Effects of Dimension-eight Operators Describing Anomalous Neutral Triple Gauge Boson Interactions at FCC-hh}},
	volume = {935},
	year = {2018}}

@article{Rauch:2016pai,
	archiveprefix = {arXiv},
	author = {Rauch, Michael},
	eprint = {1610.08420},
	journal = {Habilitation thesis, KIT},
	month = {10},
	primaryclass = {hep-ph},
	reportnumber = {KA-TP-35-2016},
	title = {{Vector-Boson Fusion and Vector-Boson Scattering}},
	year = {2016}}

@article{ATLAS:2016snd,
	archiveprefix = {arXiv},
	author = {Aaboud, Morad and others},
	collaboration = {ATLAS},
	doi = {10.1103/PhysRevD.96.012007},
	eprint = {1611.02428},
	journal = {Phys. Rev. D},
	number = {1},
	pages = {012007},
	primaryclass = {hep-ex},
	reportnumber = {CERN-EP-2016-167},
	title = {{Measurement of $W^{\pm}W^{\pm}$ vector-boson scattering and limits on anomalous quartic gauge couplings with the ATLAS detector}},
	volume = {96},
	year = {2017}}

@article{CMS:2016gct,
	archiveprefix = {arXiv},
	author = {Khachatryan, Vardan and others},
	collaboration = {CMS},
	doi = {10.1007/JHEP06(2017)106},
	eprint = {1612.09256},
	journal = {JHEP},
	pages = {106},
	primaryclass = {hep-ex},
	reportnumber = {CMS-SMP-14-011, CERN-EP-2016-289},
	title = {{Measurement of electroweak-induced production of W$\gamma$ with two jets in pp collisions at $ \sqrt{s}=8 $ TeV and constraints on anomalous quartic gauge couplings}},
	volume = {06},
	year = {2017}}

@article{ATLAS:2025yxf,
	archiveprefix = {arXiv},
	author = {Aad, Georges and others},
	collaboration = {ATLAS},
	doi = {10.1016/j.physletb.2025.140050},
	eprint = {2509.14070},
	journal = {Phys. Lett. B},
	pages = {140050},
	primaryclass = {hep-ex},
	reportnumber = {CERN-EP-2025-187},
	title = {{Observation of W+W-{\ensuremath{\gamma}} production in pp collisions at s = 13 TeV with the ATLAS detector and constraints on anomalous quartic gauge-boson couplings}},
	volume = {873},
	year = {2026}}

@article{ATLAS:2016qjc,
	archiveprefix = {arXiv},
	author = {Aad, Georges and others},
	collaboration = {ATLAS},
	doi = {10.1103/PhysRevD.93.112002},
	eprint = {1604.05232},
	journal = {Phys. Rev. D},
	number = {11},
	pages = {112002},
	primaryclass = {hep-ex},
	reportnumber = {CERN-EP-2016-049},
	title = {{Measurements of $Z\gamma$ and $Z\gamma\gamma$ production in $pp$ collisions at $\sqrt{s}=$ 8 TeV with the ATLAS detector}},
	volume = {93},
	year = {2016}}

@article{CMS:2021jji,
	archiveprefix = {arXiv},
	author = {Tumasyan, Armen and others},
	collaboration = {CMS},
	doi = {10.1007/JHEP10(2021)174},
	eprint = {2105.12780},
	journal = {JHEP},
	pages = {174},
	primaryclass = {hep-ex},
	reportnumber = {CMS-SMP-19-013, CERN-EP-2021-073},
	title = {{Measurements of the pp $\to$ W$^\pm\gamma\gamma$ and pp $\to$ Z$\gamma\gamma$ cross sections at $\sqrt{s} =$ 13 TeV and limits on anomalous quartic gauge couplings}},
	volume = {10},
	year = {2021}}

@article{ATLAS:2022wmu,
	archiveprefix = {arXiv},
	author = {Aad, Georges and others},
	collaboration = {ATLAS},
	doi = {10.1140/epjc/s10052-023-11579-8},
	eprint = {2211.14171},
	journal = {Eur. Phys. J. C},
	number = {6},
	pages = {539},
	primaryclass = {hep-ex},
	reportnumber = {CERN-EP-2022-192},
	title = {{Measurement of $Z\gamma \gamma $ production in pp collisions at $\sqrt{s}= 13$~TeV with the ATLAS detector}},
	volume = {83},
	year = {2023}}

@article{CMS:2017fhs,
	archiveprefix = {arXiv},
	author = {Sirunyan, Albert M and others},
	collaboration = {CMS},
	doi = {10.1103/PhysRevLett.120.081801},
	eprint = {1709.05822},
	journal = {Phys. Rev. Lett.},
	number = {8},
	pages = {081801},
	primaryclass = {hep-ex},
	reportnumber = {CMS-SMP-17-004, CERN-EP-2017-232},
	title = {{Observation of electroweak production of same-sign W boson pairs in the two jet and two same-sign lepton final state in proton-proton collisions at $\sqrt{s} = $ 13 TeV}},
	volume = {120},
	year = {2018}}

@article{ATLAS:2019cbr,
	archiveprefix = {arXiv},
	author = {Aaboud, Morad and others},
	collaboration = {ATLAS},
	doi = {10.1103/PhysRevLett.123.161801},
	eprint = {1906.03203},
	journal = {Phys. Rev. Lett.},
	number = {16},
	pages = {161801},
	primaryclass = {hep-ex},
	reportnumber = {CERN-EP-2019-008},
	title = {{Observation of electroweak production of a same-sign $W$ boson pair in association with two jets in $pp$ collisions at $\sqrt{s}=13$ TeV with the ATLAS detector}},
	volume = {123},
	year = {2019}}

@article{Yilmaz:2019cue,
	archiveprefix = {arXiv},
	author = {Yilmaz, A. and Senol, A. and Denizli, H. and Turk Cakir, I. and Cakir, O.},
	doi = {10.1140/epjc/s10052-020-7731-2},
	eprint = {1906.03911},
	journal = {Eur. Phys. J. C},
	number = {2},
	pages = {173},
	primaryclass = {hep-ph},
	title = {{Sensitivity on Anomalous Neutral Triple Gauge Couplings via $ZZ$ Production at FCC-hh}},
	volume = {80},
	year = {2020}}

@article{Senol:2019qyl,
	archiveprefix = {arXiv},
	author = {Senol, A. and Denizli, H. and Yilmaz, A. and Turk Cakir, I. and Cakir, O.},
	doi = {10.1016/j.physletb.2020.135255},
	eprint = {1910.03843},
	journal = {Phys. Lett. B},
	pages = {135255},
	primaryclass = {hep-ph},
	title = {{The projections on $ZZ\gamma$ and $Z\gamma\gamma$ couplings via $\nu\bar \nu \gamma$ production in HL-LHC and HE-LHC}},
	volume = {802},
	year = {2020}}

@article{Yilmaz:2021ule,
	archiveprefix = {arXiv},
	author = {Yilmaz, Ali},
	doi = {10.1016/j.nuclphysb.2021.115471},
	eprint = {2102.01989},
	journal = {Nucl. Phys. B},
	pages = {115471},
	primaryclass = {hep-ph},
	title = {{Search for the limits on anomalous neutral triple gauge couplings via ZZ production in the $\ell\ell\nu\nu$ channel at FCC-hh}},
	volume = {969},
	year = {2021}}

@article{Senol:2019swu,
	archiveprefix = {arXiv},
	author = {Senol, A. and Denizli, H. and Yilmaz, A. and Turk Cakir, I. and Cakir, O.},
	doi = {10.5506/APhysPolB.50.1597},
	eprint = {1906.04589},
	journal = {Acta Phys. Polon. B},
	pages = {1597},
	primaryclass = {hep-ph},
	title = {{Study on Anomalous Neutral Triple Gauge Boson Couplings from Dimension-eight Operators at the HL-LHC}},
	volume = {50},
	year = {2019}}

@article{Covarelli:2021gyz,
	archiveprefix = {arXiv},
	author = {Covarelli, Roberto and Pellen, Mathieu and Zaro, Marco},
	doi = {10.1142/S0217751X2130009X},
	eprint = {2102.10991},
	journal = {Int. J. Mod. Phys. A},
	number = {16},
	pages = {2130009},
	primaryclass = {hep-ph},
	reportnumber = {FR-PHENO-2021-05, TIF-UNIMI-2021-2, VBSCAN-PUB-02-21},
	title = {{Vector-Boson scattering at the LHC: Unraveling the electroweak sector}},
	volume = {36},
	year = {2021}}

@article{Alloul:2013bka,
	archiveprefix = {arXiv},
	author = {Alloul, Adam and Christensen, Neil D. and Degrande, C{\'e}line and Duhr, Claude and Fuks, Benjamin},
	doi = {10.1016/j.cpc.2014.04.012},
	eprint = {1310.1921},
	journal = {Comput. Phys. Commun.},
	pages = {2250--2300},
	primaryclass = {hep-ph},
	reportnumber = {CERN-PH-TH-2013-239, MCNET-13-14, IPPP-13-71, DCPT-13-142, PITT-PACC-1308},
	title = {{FeynRules 2.0 - A complete toolbox for tree-level phenomenology}},
	volume = {185},
	year = {2014}}

@article{Degrande:2011ua,
	archiveprefix = {arXiv},
	author = {Degrande, Celine and Duhr, Claude and Fuks, Benjamin and Grellscheid, David and Mattelaer, Olivier and Reiter, Thomas},
	doi = {10.1016/j.cpc.2012.01.022},
	eprint = {1108.2040},
	journal = {Comput. Phys. Commun.},
	pages = {1201--1214},
	primaryclass = {hep-ph},
	reportnumber = {CP3-11-25, IPHC-PHENO-11-04, IPPP-11-39, DCPT-11-78, MPP-2011-68},
	title = {{UFO - The Universal FeynRules Output}},
	volume = {183},
	year = {2012}}

@article{Ball:2012cx,
	archiveprefix = {arXiv},
	author = {Ball, Richard D. and others},
	doi = {10.1016/j.nuclphysb.2012.10.003},
	eprint = {1207.1303},
	journal = {Nucl. Phys. B},
	pages = {244--289},
	primaryclass = {hep-ph},
	reportnumber = {EDINBURGH-2012-08, IFUM-FT-997, FR-PHENO-2012-014, RWTH-TTK-12-25, CERN-PH-TH-2012-037, SFB-CPP-12-47},
	title = {{Parton distributions with LHC data}},
	volume = {867},
	year = {2013}}

@article{Read:2002hq,
	author = {Read, Alexander L.},
	doi = {10.1088/0954-3899/28/10/313},
	editor = {Whalley, M. R. and Lyons, L.},
	journal = {J. Phys. G},
	pages = {2693--2704},
	title = {{Presentation of search results: The $CL_s$ technique}},
	volume = {28},
	year = {2002}}

@article{Brivio:2017vri,
	archiveprefix = {arXiv},
	author = {Brivio, Ilaria and Trott, Michael},
	doi = {10.1016/j.physrep.2018.11.002},
	eprint = {1706.08945},
	journal = {Phys. Rept.},
	pages = {1--98},
	primaryclass = {hep-ph},
	title = {{The Standard Model as an Effective Field Theory}},
	volume = {793},
	year = {2019}}

@article{Cowan:2010js,
	archiveprefix = {arXiv},
	author = {Cowan, Glen and Cranmer, Kyle and Gross, Eilam and Vitells, Ofer},
	doi = {10.1140/epjc/s10052-011-1554-0},
	eprint = {1007.1727},
	journal = {Eur. Phys. J. C},
	note = {[Erratum: Eur.Phys.J.C 73, 2501 (2013)]},
	pages = {1554},
	primaryclass = {physics.data-an},
	title = {{Asymptotic formulae for likelihood-based tests of new physics}},
	volume = {71},
	year = {2011}}

@article{Bozzi:2009ig,
	archiveprefix = {arXiv},
	author = {Bozzi, G. and Campanario, F. and Hankele, V. and Zeppenfeld, D.},
	doi = {10.1103/PhysRevD.81.094030},
	eprint = {0911.0438},
	journal = {Phys. Rev. D},
	pages = {094030},
	primaryclass = {hep-ph},
	reportnumber = {KA-TP-11-2009, SFB-CPP-09-105, FTUV-09-1101, IFUM-938-FT},
	title = {{NLO QCD corrections to W+W- gamma and Z Z gamma production with leptonic decays}},
	volume = {81},
	year = {2010}}

@article{Alwall:2014hca,
	archiveprefix = {arXiv},
	author = {Alwall, J. and Frederix, R. and Frixione, S. and Hirschi, V. and Maltoni, F. and Mattelaer, O. and Shao, H. -S. and Stelzer, T. and Torrielli, P. and Zaro, M.},
	doi = {10.1007/JHEP07(2014)079},
	eprint = {1405.0301},
	journal = {JHEP},
	pages = {079},
	primaryclass = {hep-ph},
	reportnumber = {CERN-PH-TH-2014-064, CP3-14-18, LPN14-066, MCNET-14-09, ZU-TH-14-14},
	title = {{The automated computation of tree-level and next-to-leading order differential cross sections, and their matching to parton shower simulations}},
	volume = {07},
	year = {2014}}

@article{Sjostrand:2014zea,
	archiveprefix = {arXiv},
	author = {Sj{\"o}strand, Torbj{\"o}rn and Ask, Stefan and Christiansen, Jesper R. and Corke, Richard and Desai, Nishita and Ilten, Philip and Mrenna, Stephen and Prestel, Stefan and Rasmussen, Christine O. and Skands, Peter Z.},
	doi = {10.1016/j.cpc.2015.01.024},
	eprint = {1410.3012},
	journal = {Comput. Phys. Commun.},
	pages = {159--177},
	primaryclass = {hep-ph},
	reportnumber = {LU-TP-14-36, MCNET-14-22, CERN-PH-TH-2014-190, FERMILAB-PUB-14-316-CD, DESY-14-178, SLAC-PUB-16122},
	title = {{An introduction to PYTHIA 8.2}},
	volume = {191},
	year = {2015}}

@article{deFavereau:2013fsa,
	archiveprefix = {arXiv},
	author = {de Favereau, J. and Delaere, C. and Demin, P. and Giammanco, A. and Lema{\^\i}tre, V. and Mertens, A. and Selvaggi, M.},
	collaboration = {DELPHES 3},
	doi = {10.1007/JHEP02(2014)057},
	eprint = {1307.6346},
	journal = {JHEP},
	pages = {057},
	primaryclass = {hep-ex},
	title = {{DELPHES 3, A modular framework for fast simulation of a generic collider experiment}},
	volume = {02},
	year = {2014}}

@article{TMVA:2007ngy,
	archiveprefix = {arXiv},
	author = {Hocker, Andreas and others},
	collaboration = {TMVA},
	eprint = {physics/0703039},
	journal = {PoS},
	pages = {040},
	reportnumber = {CERN-OPEN-2007-007},
	title = {{TMVA - Toolkit for Multivariate Data Analysis}},
	volume = {ACAT},
	year = {2007}}

@article{Unel:2021edl,
	archiveprefix = {arXiv},
	author = {Unel, G. and Sekmen, S. and Toon, A. M. and Gokturk, B. and Orgen, B. and Paul, A. and Ravel, N. and Setpal, J.},
	doi = {10.3389/fdata.2021.659986},
	eprint = {2101.09031},
	journal = {Front. Big Data},
	pages = {659986},
	primaryclass = {hep-ph},
	title = {{CutLang v2: Advances in a Runtime-Interpreted Analysis Description Language for HEP Data}},
	volume = {4},
	year = {2021}}

@article{Arnold:2008rz,
	archiveprefix = {arXiv},
	author = {Arnold, K. and others},
	doi = {10.1016/j.cpc.2009.03.006},
	eprint = {0811.4559},
	journal = {Comput. Phys. Commun.},
	pages = {1661--1670},
	primaryclass = {hep-ph},
	reportnumber = {KA-TP-31-2008, SFB-CPP-08-95},
	title = {{VBFNLO: A Parton level Monte Carlo for processes with electroweak bosons}},
	volume = {180},
	year = {2009}}

@article{Baglio:2014uba,
	archiveprefix = {arXiv},
	author = {Baglio, J. and others},
	eprint = {1404.3940},
	month = {4},
	primaryclass = {hep-ph},
	reportnumber = {FTUV-14-2903, IFIC-14-26, KA-TP-10-2014, LPN14-062, MAN-HEP-2014-03},
	title = {{Release Note - VBFNLO 2.7.0}},
	year = {2014}}

\end{document}